\documentclass[fleqn,usenatbib]{mnras}
\usepackage{newtxtext,newtxmath}
\usepackage[T1]{fontenc}
\usepackage{threeparttable}
\usepackage{longtable}
\DeclareRobustCommand{\VAN}[3]{#2}
\let\VANthebibliography\thebibliography
\def\thebibliography{\DeclareRobustCommand{\VAN}[3]{##3}\VANthebibliography}

\usepackage{graphicx}	
\usepackage{amsmath}	
\usepackage{amssymb}	

\usepackage{caption}
\title[Heavy Element Abundances in the GC]{{\it Chandra} X-ray Measurement of Heavy Element Abundances of Wolf-Rayet Stars in the Galactic Center}

\author[Z.Q. Hua et al.]{
Ziqian Hua,$^{1,2}$\thanks{E-mail: zqhua@smail.nju.edu.cn}
Zhiyuan Li,$^{1,2,3}$\thanks{E-mail: lizy@nju.edu.cn}
\\
$^{1}$School of Astronomy and Space Science, Nanjing University, Nanjing 210023, China\\
$^{2}$Key Laboratory of Modern Astronomy and Astrophysics (Nanjing University), Ministry of Education, Nanjing 210023, China\\
$^{3}$Institute of Science and Technology for Deep Space Exploration, Suzhou Campus, Nanjing University, Suzhou 215163, China\\
}

\date{Accepted XXX. Received YYY; in original form ZZZ}

\begin{document}
\label{firstpage}
\pagerange{\pageref{firstpage}--\pageref{lastpage}}
\maketitle

\begin{abstract}
Elemental abundances hold important information about the star formation history in the Galactic Center.
The thermal X-ray spectra of certain stars can provide a robust probe of elemental abundances, mainly through the presence of K-shell emission lines.
In this work, based on deep archival {\it Chandra} observations, we obtain X-ray measurements of five heavy elements (Si, S, Ar, Ca and Fe) for three sources in the Arches cluster, one source in the Quintuplet cluster, as well as a field source known as Edd 1, which are all probable WR stars exhibiting a high quality X-ray spectrum.  
A two-temperature, non-equilibrium ionization plasma model is employed for the spectral fit, taking into account light element compositions characteristic of WR star winds, which is substantially depleted in hydrogen but enriched in nitrogen and/or carbon. 
It is found that the Arches and Quintuplet WR stars share similar abundances of Si, S, and Ar, while exhibiting distinct Ca and Fe abundances, which may be understood as due to dust depletion of the latter two elements in Quintuplet. 
The observed near-solar or sub-solar metallicity of the WR star winds can
be naturally understood as the result of nucleosynthesis and internal mixing of the parent star, which have a supersolar initial metallicity as expected for the Galactic center in general.
Implications of our findings on the origin of the young star clusters and isolated massive stars in the Galactic center, as well as the elemental composition of the accretion flow onto Sgr A*, are addressed. 
\end{abstract}

\begin{keywords}
Galaxy: centre -- Galaxy: abundances -- stars: Wolf-Rayet -- X-rays: ISM.
\end{keywords}



\section{Introduction}
\label{sec:Intro}
In modern astrophysics, the Galactic Center (GC) is an important laboratory by virtue of its proximity ($D \approx8~{\rm kpc}$, \citealp{2019ApJ...882L..27D, 2020A&A...636L...5G}).
The star formation history in the GC, including the rate, efficiency and distribution of newly formed stars, offers crucial insights into the unique role of the nuclear environment centered on a supermassive black hole (SMBH), known as Sgr A*.
High-resolution multiwavelength observations have provided valuable information about the dense gas and stellar components in the GC.
Surrounding Sgr A*, a total mass of $\sim 2.5\times10^7~{\rm M_{\odot}}$ stellar objects constitute the dense, spheroidal nuclear stellar cluster (NSC), which dominates the starlight of the innermost $\sim$20 pc \citep{2010RvMP...82.3121G,2017MNRAS.466.4040F}.
A larger, more flattened structure, called the nuclear stellar disk (NSD), occupies the outer GC region up to $\sim$300 pc and contains a stellar mass of $\sim 1\times10^9~{\rm M_{\odot}}$ \citep{2002A&A...384..112L,2022MNRAS.512.1857S}.
Co-spatial with the NSD, the ring-like central molecular zone (CMZ; \citealp{1996ARA&A..34..645M}) contains $\sim (3-5)\times10^7~{\rm M_{\odot}}$ of dense molecular gas \citep{1998A&A...331..959D,2007A&A...467..611F} and hosts most present-day star formation in the GC \citep{2009ApJ...702..178Y}.
The molecular clouds and stellar objects exhibit similar rotation velocities \citep{2015ApJ...812L..21S,2021A&A...650A.191S}, and there is evidence that the star formation in the CMZ contributes to the build-up of the NSD \citep{2020MNRAS.492.4500B}.
The complex interplay between gravity and various baryonic processes governs star formation in the GC.

In recent years, many studies have explored the elemental abundances in the GC.
It is found that the NSC predominantly comprises metal-rich stars with a median metallicity of $\rm{[M/H]}\approx-0.16$, spanning from $\sim-0.3$ to 0.3. 
This suggests a complex formation history involving multiple star formation episodes and potential mergers \citep{2017AJ....154..239R}.
Meanwhile, the NSD is characterized by a dynamically cool and metal-rich stellar population.
\citet{2021A&A...650A.191S} found that the NSD's metallicity distribution differs from that of the NSC and the inner bulge, with a higher fraction of metal-rich stars.
This distinction implies that the NSD may have formed from gas in the CMZ, leading to a unique chemical and kinematic profile. 
Investigation into the transition region between the NSC and NSD reveals a steep metallicity gradient, decreasing from the NSC out to the NSD \citep{2022MNRAS.513.5920F}.

The GC hosts numerous young, massive stars, forming under an extreme condition of strong tidal force, intense radiation and high pressure. 
These stars generate ultraviolet radiation and stellar winds, injecting energy into and shaping the interstellar medium (ISM). 
A significant fraction of these massive, early-type stars are concentrated in three star clusters: the young nuclear cluster (YNC, which is part of the NSC but occupies the central parsec; \citealp{1991ApJ...382L..19K}), the Arches cluster \citep{1995AJ....109.1676N}, and the Quintuplet cluster \citep{1990ApJ...351...83N}.
These three star clusters are believed to be formed within a few million years and contain numerous massive stars \citep{2019ApJS..244...35L}. 
The YNC contains 33 currently known Wolf-Rayet (WR) stars (according to the Galactic Wolf Rayet Catalog\footnote{http://pacrowther.staff.shef.ac.uk/WRcat/index.php}) and more than 100 main-sequence massive stars with an age of 4-6 Myr \citep{2010RvMP...82.3121G,2013ApJ...764..155L}.
There are 13 WR stars detected in the Arches cluster and 21 WR stars in the Quintuplet cluster \citep{2009A&A...501..563E,2012A&A...540A..14L}.
Estimation of their ages is provided by \citet{2002ApJ...581..258F} and \citet{1999ApJ...514..202F}, with 2-3 Myr for Arches and 3-4 Myr for Quintuplet.
Accordingly, more WN-type stars exist in Arches, which are younger and have a larger fraction of hydrogen preserved in their winds.
In addition to the WR stars in the young massive clusters, a comparable number of isolated WR stars are also found distributed throughout the GC \citep{1996A&A...315L.193V}.
How these isolated WR stars form and evolve remains an open question. They might have been born in  massive clusters that have dissolved under strong tidal force, or they could have been formed in genuine isolation. 

When these massive stars reside in binary systems, the intense stellar winds will collide with each other, forming the so-called colliding-wind binaries (CWBs).
The colliding wind generates strong shocks and produces a hot plasma (with a temperature $\gtrsim10^6$ K). Relativistic particles are also observed in such systems \citep{1993ApJ...402..271E,1995Ap&SS.224..367S}, which can be accelerated through diffusive shock acceleration \citep{1983RPPh...46..973D}.
These systems can be bright in the X-ray band. X-ray emission has been detected in several CWBs including Apep \citep{2023A&A...672A.109D}, $\eta$ Car \citep{2018NatAs...2..731H} and $\gamma$ Vel \citep{1995A&A...298..549W}.
The overall X-ray spectral shape of typical CWBs contains a soft, thermal component as well as a hard, non-thermal component.
The latter is due to synchrotron or inverse Compton scattering of relativistic electrons, which contributes to hard X-rays and even $\gamma$-rays.
The wind velocity and the post-shock temperature determine the thermal component, which contains a bremsstrahlung continuum and  emission lines mainly from the K-shell transitions of various elements.
This provides an important probe of the elemental abundances in the wind material of these systems.

In a recent study, \citet{2023MNRAS.522..635H} selected four prominent diffuse X-ray features in the NSC to investigate the gas-phase heavy element abundances.
These diffuse X-ray features presumably originate from shock-heated stellar winds from the aforementioned $\sim30$ WR stars in the YNC. 
In particular, the most compact one of these features is coincident with the infrared source IRS 13E, which is shown to be the result of the colliding winds of two WR stars \citep{2020ApJ...897..135Z}.
\citet{2023MNRAS.522..635H} applied a two-temperature, non-equilibrium spectral model to the high quality X-ray spectra from 5.7 Ms {\it Chandra} observations and made assumptions about the compositions of light elements (in particular, a low hydrogen content) dominating the bremsstrahlung emission, which are based on infrared spectroscopy \citep{2007A&A...468..233M}.
These authors determined that the abundances of Si, S, Ar, Ca and Fe of the diffuse X-ray features are generally subsolar, but with a supersolar $\alpha$/Fe abundance ratio.

Outside the NSC, the WR stars in Arches and Quintuplet as well as in the field can in principle provide a useful diagnostic of the heavy element abundances across the GC, helping with a deeper understanding of the most recent star formation episode in this unique environment. 
We carry out such an effort in this work, utilizing recently available deep ${\it Chandra}$ observations.
The paper is organized as follows. 
In Section \ref{sec:data}, we describe the ${\it Chandra}$ observations and the data reduction procedure, followed by details about the target selection. 
In Section \ref{sec:spec}, we outline the spectral analysis process and provide the measurements of the heavy element abundances.
In Section \ref{sec:Discussion}, we discuss potential caveats in the measurements, indications on the initial metallicity of the three star clusters, and implications on the GC environment.
A summary of this study is provided in Section~\ref{sec:sum}.
Throughout this work, we adopt a distance of 8 kpc (\citealp{2019ApJ...882L..27D, 2020A&A...636L...5G}) for the Galactic center.

\begin{figure*}
	\includegraphics[width= \textwidth]{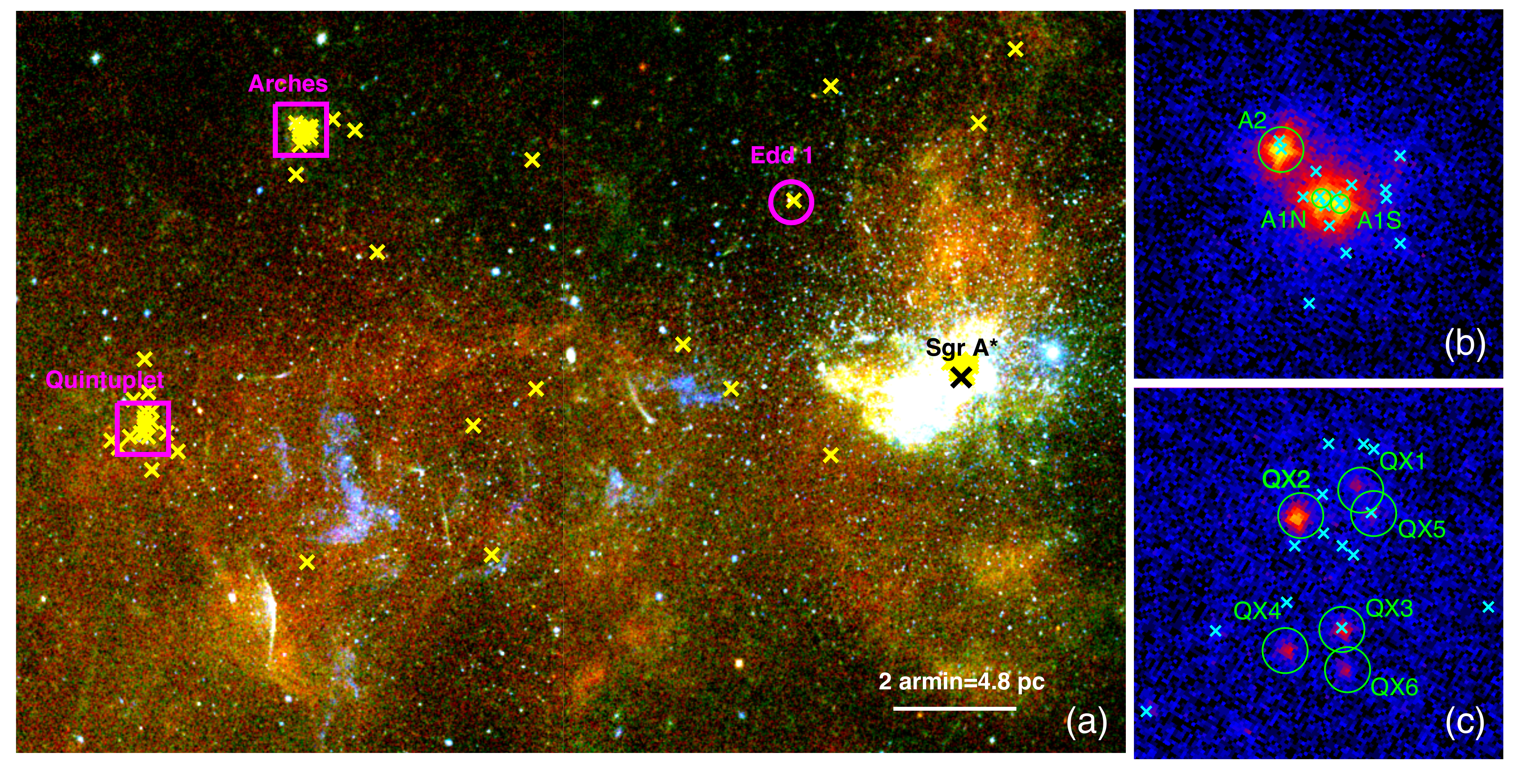}
 \caption{(a) Tricolor image of flux map in the GC region (red: 2--3.3 keV, green: 3.3--4.7 keV, blue: 4.7--8 keV). The image size is $0.3^{\circ}\times0.2^{\circ}$ and the horizontal direction is parallel to the Galactic plane. The intensity in each band has been corrected for effective exposure and smoothed with a Gaussian kernel of 2 pixels. Sgr A* is labeled with the black cross, while the locations of known WR stars \citep{2001NewAR..45..135V,2006A&A...458..453V,2009A&A...494.1137L} are marked by yellow crosses. The two young massive clusters, Arches and Quintuplet, are indicated by magenta boxes. The isolated WR binary Edd 1 is marked by the magenta circle. (b) Zoom-in view of the Arches cluster in the 2--8 keV energy range, combining 54 individual observations. (c) Zoom-in view of the Quintuplet cluster in the 2--8 keV energy range, combining 59 individual observations.
 The size of the two zoom-in panels is $50\arcsec\times50\arcsec$.
 Locations of known WR stars are marked by crosses, while the source extraction regions are marked by circles.}
    \label{fig:xray_image}
\end{figure*}

\section{Data preparation and target selection}
\label{sec:data}
\subsection{{\it Chandra} Data}
The GC region has been extensively observed by {\it Chandra} for over 20 years since its deployment, with a 
focus on Sgr A* and its immediate surroundings.
Many of these observations also encompassed the Arches and Quintuplet clusters, providing a valuable dataset for analyzing their X-ray properties.
This study utilizes a total of 60 {\it Chandra} Advanced CCD Imaging Spectrometer (ACIS) observations that cover Arches and/or Quintuplet, including the 100 ks observation (ObsID 4500) previously used by \citet{2006MNRAS.371...38W}, 43 observations taken in 2022 chiefly to monitor the X-ray reflection from dense molecular clouds near Sgr A*, and a handful of complementary exposures.
The detailed information of the adopted observations are given in Table~\ref{tab:obs_info}.
The total effective exposure, after filtering bad time intervals, amounts to 1.44 Ms for the Arches cluster and 1.67 Ms for the Quintuplet cluster. 

Thanks to its proximity to Sgr A*, Edd 1 (see Section~\ref{subsubsec:Edd1} below), located at ${\rm RA=17h45m36s.12,~Dec=-28^{\circ}56'38.7''}$, is covered by a large number of {\it Chandra} observations targeting the NSC.
We utilize 48 ACIS-I observations, 21 ACIS-G{\footnote{Observations taken
in 2012, with the High Energy Transmission Grating in operation
and the S3 CCD on-axis. Only the zeroth order image data of these observations are used \citep{2023MNRAS.522..635H}.} observations and 14 ACIS-S observations for Edd 1, resulting in a cumulative exposure time of 1.5 Ms, 1.9 Ms, and 0.6 Ms, respectively. 
All these observations were previously adopted by 
\citet{2023MNRAS.522..635H} to measure the heavy element abundances of diffuse X-ray features in the NSC.

All relevant {\it Chandra} data were downloaded from the public archive and uniformly reprocessed with CIAO 4.14 and CALDB 4.9.2, following the standard procedures detailed in \citet{2020ApJ...897..135Z}. A tricolor image of the flux map in the GC region is shown in Fig.~\ref{fig:xray_image}(a), indicating the locations of Arches and Quintuplet, as well as Edd 1. The locations of the known WR stars from \citet{2001NewAR..45..135V,2006A&A...458..453V} are also marked by yellow crosses.
Figures~\ref{fig:xray_image}(b) and (c) further display a zoom-in view of the two clusters in the energy range of 2--8 keV. 
Photons with an energy below $\sim2$ keV from the GC region are strongly absorbed.   

\subsection{Target Selection}
\subsubsection{Arches}
\citet{2006MNRAS.371...38W} reported a deep {\it Chandra} ACIS-I observation of the Arches and Quintuplet clusters.
Three bright point sources, named A1N, A1S and A2, were identified in the core region of the Arches cluster (Fig.~\ref{fig:xray_image}b). 
Their near-infrared counterparts have been classified as late-WN types and no clear sign indicating binarity has been found.
These three prominent X-ray sources are surrounded by diffuse X-ray emission.
While the three point-like sources clearly originate from CWBs, the diffuse emission is more complex, involving the cluster wind and its interaction with the surrounding dense molecular cloud.
Here we focus on the three sources and analyze their spectra separately.
The two southern sources, A1N and A1S, are adjacent to each other.
To minimize possible mutual contamination between them, we use 1.25$\arcsec$-radius circular regions to extract their spectra. 
The northern source, A2, has no bright neighbors, so we adopt a 3$\arcsec$-radius circular region, which is roughly the size of the 90\% enclosed energy circle at this position.
The background region is taken to be a ring with inner and outer radii of $10\arcsec$ and $20\arcsec$, with any enclosed point sources excluded.
We inspected and found only mild inter-observation flux variation in all three sources, with no evidence for periodicity. Hence we will consider a single co-added spectrum for all three Arches sources.

\subsubsection{Quintuplet}
The Quintuplet cluster is less luminous in X-rays, containing a few point-like sources as well as diffuse emission (Fig.~\ref{fig:xray_image}c). 
Compared with \citet{2006MNRAS.371...38W}, the new {\it Chandra} observations provide significantly more information and new insights into the point sources and diffuse emission in this cluster.
As shown in Fig.~\ref{fig:xray_image}(c), we can clearly distinguish the sources QX1-QX5 discussed in \citet{2006MNRAS.371...38W}, and are also able to identify a new source which is named QX6.
The X-ray spectral properties of the Quintuplet sources vary significantly. 
\citet{2006MNRAS.371...38W} claim that QX1, with a soft spectrum, is a foreground source, while QX4, with no identified near-infrared counterpart, is either a background AGN or a highly obscured object.
We find that QX6 is likely a cataclysmic variable due to a hard continuum and the presence of 6.4 and 6.7 keV Fe lines (see Appendix~\ref{QX6}).
QX2, QX3 and QX5 have spectral properties consistent with CWB systems, but the latter two have only a moderate signal-to-noise ratio (S/N) due to a low flux.
Therefore, we only measure the abundances of QX2, which we hypothesize as a WR star according to its X-ray spectral characteristics and take it as a representative of the Quintuplet cluster.
No significant flux variability is found for QX2.
A circular region with 3$\arcsec$-radius is used to extract the source spectrum. The background region is taken to be a ring with inner and outer radii of $20\arcsec$ and $40\arcsec$, with any enclosed point sources excluded. 
Spectral fit results are insensitive to the exact choice of the background region thanks to the high S/N of QX2. 
We also found only mild inter-observation flux variation in QX2, with no evidence for periodicity. Likewise we consider a single co-added spectrum for this source.

\subsubsection{Edd 1}
\label{subsubsec:Edd1}
\begin{figure}
    \includegraphics[width= 0.48\textwidth]{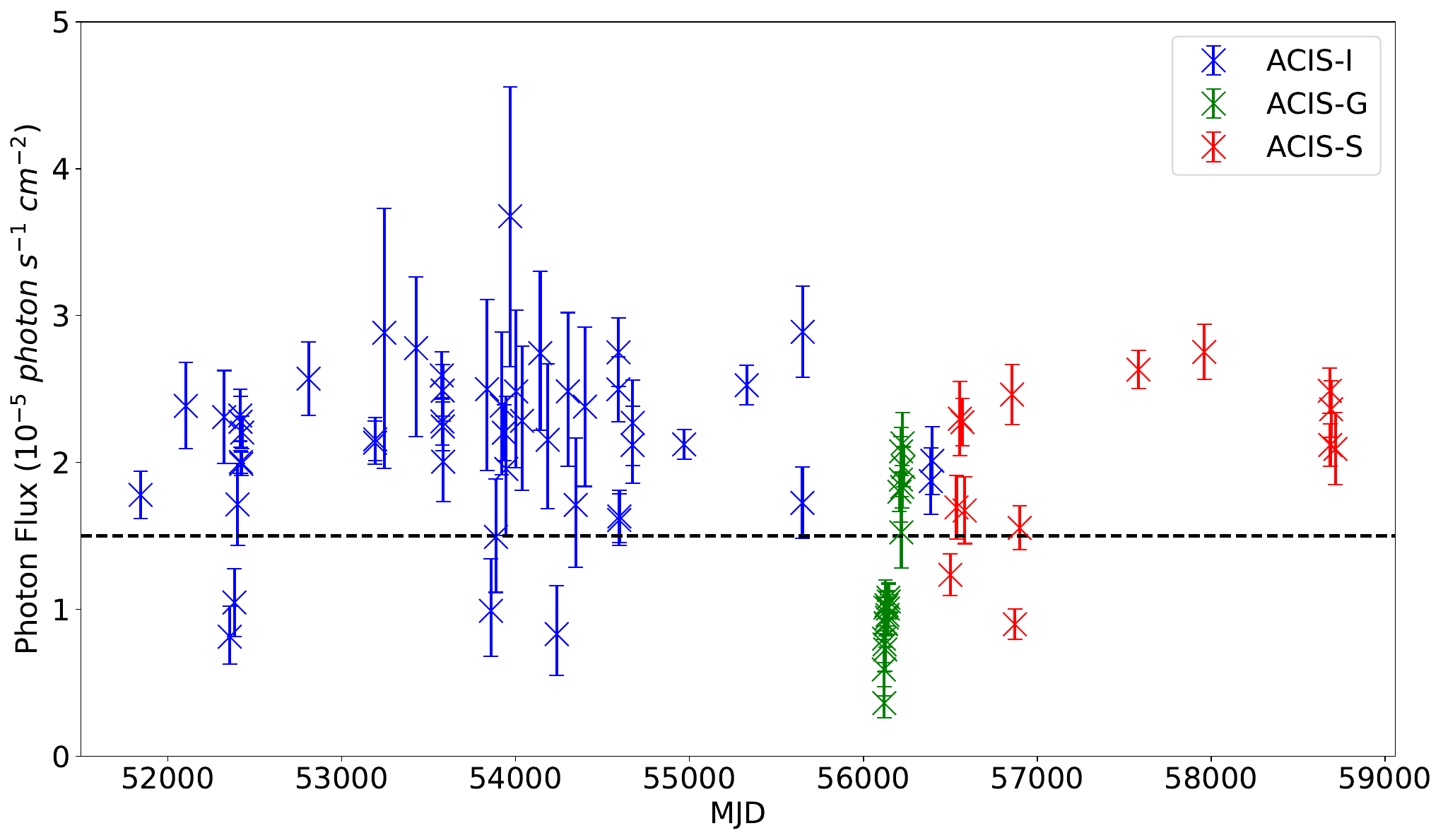}
    \caption{The observed 2--8 keV photon flux of Edd 1 in different observations. The flux and corresponding 1-$\sigma$ errors are calculated from within the 75$\%$ enclosed-energy radius with CIAO tool ${\it aprates}$. Observations using ACIS-I, ACIS-G and ACIS-S are labelled with blue, green and red crosses, respectively. The dotted line indicates a photon flux level of $1.5\times10^{-5}~{\rm ph~cm^{-2}~s^{-1}}$, below (above) which the low (high) state is defined.}
    \label{fig:edd1_lc}
\end{figure}

Some of the isolated WR stars in the GC are also likely to reside in binary systems, thus manifesting themselves as prominent X-ray sources.
Eight X-ray sources detected by \citet{2009ApJS..181..110M} and \citet{2018ApJS..235...26Z} are positionally coincident with known WR stars \citep{2001NewAR..45..135V,2006A&A...458..453V} and exhibit spectral characteristics similar to CWB systems.
However, most of these sources are faint and we have limited knowledge about their stellar wind compositions.
Among them, the brightest one is CXOGC J174536.1-285638, also known as Edd 1.
\citet{2006ApJ...651..408M} confirmed the infrared counterpart of Edd 1 with 2MASS survey and found strong Bracket $\gamma$ and He I lines in its infrared spectrum.
The presence of C III, N II and He II lines indicates that Edd 1 is a binary system, which is further supported by the $189\pm6$-day X-ray period found by \citet{2008ApJ...689.1222M}.
\citet{2009A&A...507.1567C} analyzed the near-infrared spectrum of Edd 1 and identified it as a WN9h+O VIII system.
Considering the large off-axis distance and varied roll angle of Edd 1 appearing in the {\it Chandra} observations, we use the 75\% enclosed energy radius to extract its spectra, similar to the choice of \citet{2018ApJS..235...26Z} in order to minimize potential contamination from neighboring sources.
The background region is taken to be a ring with inner and outer radii of 2 and 4 times the source region radius.
In view of a $189\pm6$ day X-ray period reported by \citet{2008ApJ...689.1222M}, we generate the long-term light curve of Edd 1 using the CIAO tool ${\it aprates}$, 
correcting for the different instrumental responses.
As shown in Fig.~\ref{fig:edd1_lc}, the light curve exhibits significant flux variation, likely due to varying levels of self-absorption by stellar wind materials, following the binary orbital motion.
Since the X-ray spectral properties may vary with the orbital phase, we divide all the observations into two groups based on the observed photon flux.
The low-state (high-state) group is defined by a photon flux lower (higher) than $1.5\times10^{-5}~{\rm ph\ cm^{-2}\ s^{-1}}$, which is roughly the mean photon flux of Edd~1. 
It is evident that Edd 1 predominantly exhibits a high state in the majority of observations, while the low state is almost exclusively found during the ACIS-G observations, with the ACIS-I and ACIS-S observations giving a shallow exposure insufficient to provide high-quality spectra.
Therefore, we only utilize 12 ACIS-G observations to extract the low-state spectrum, while the high-state spectra are extracted for all three instrument sets (ACIS-I, -G and -S).

\begin{figure*}
    \includegraphics[width=\textwidth]{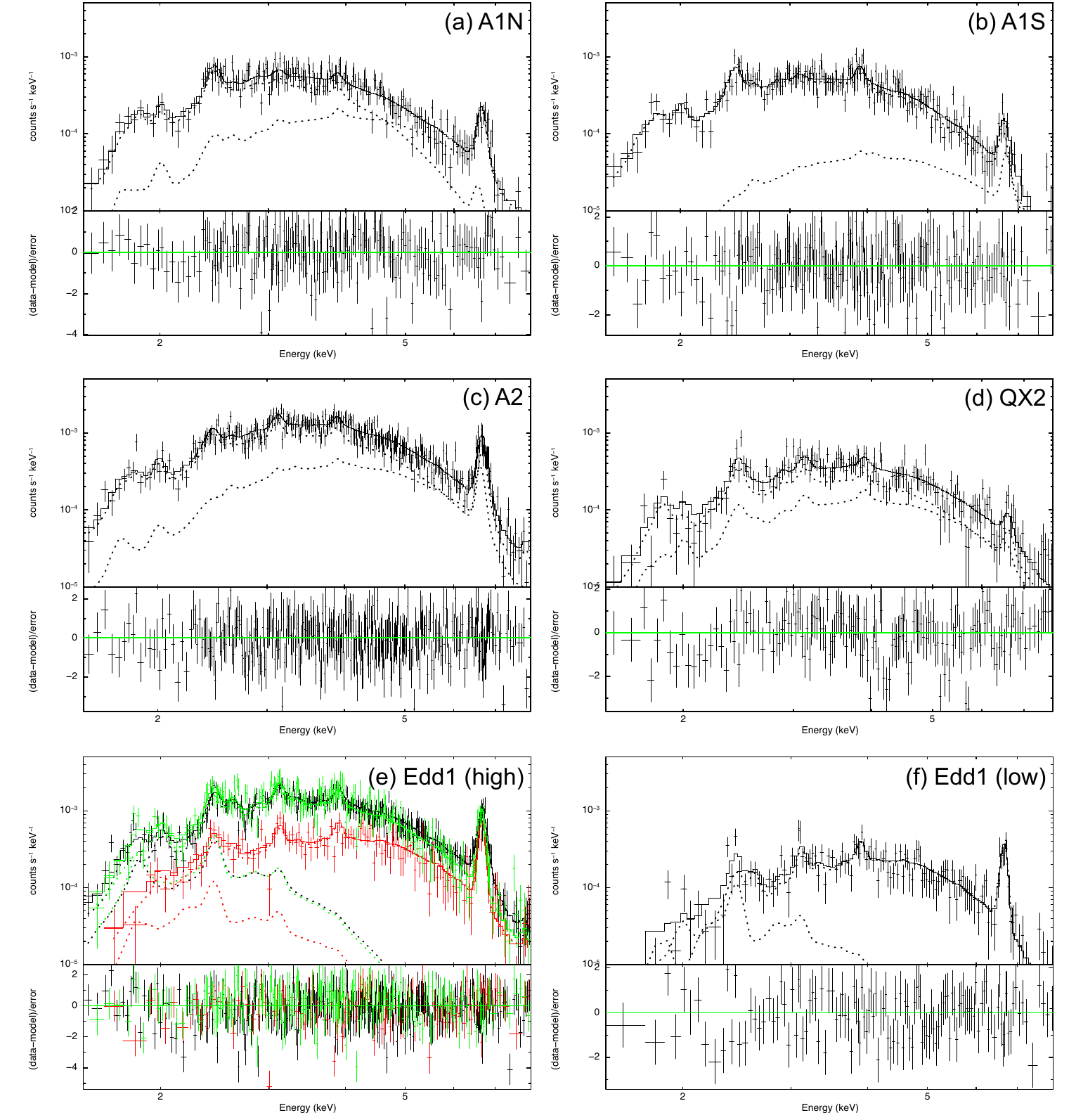}
    \caption{The background-subtracted spectra of the WR stars in the GC. (a)-(c): sources A1N, A1S and A2 in Arches; (d) source QX2 in Quintuplet; (e)-(f) Edd 1 in high- and low-state. 
    All the spectra are extracted from ACIS-I observations expect for the Edd1 high-state, where spectra are extracted from ACIS-I, -G and -S observations and shown in black, red and green respectively.
    An adaptive binning has been applied to achieve at least 10 counts and S/N greater than 3 in each spectral bin. 
    The best-fit hydrogen-depleted two-temperature NEI model, ${\it TBABS*DUSTSCAT*(VNEI+VNEI)}$ in XSPEC, is shown by solid lines, with the two NEI components shown by the dotted lines.
    The bottom panels show the relative residuals. The error bars are at the 1$\sigma$ level.}
    \label{fig:spectra}
\end{figure*}

\section{X-ray Spectral Analysis}
\label{sec:spec}
Using the CIAO tool {\it specextract}, we extracted the source spectra from individual observations and co-added them according to the corresponding instruments. 
Since the substantial degradation of the ACIS effective area occurs primarily for photon energies below $\sim2$ keV, no significant bias is expected in the co-added spectra of these GC sources, where only the emission $\gtrsim$2 keV is transparent.
Background-subtracted spectra from A1N, A1S and A2 as well as QX2 are presented in Fig.~\ref{fig:spectra}(a)-(d).
The three high-state spectra of Edd 1 is shown in Fig.~\ref{fig:spectra}(e) and the one low-state spectrum in Fig.~\ref{fig:spectra}(f).
These spectra are typical of CWB systems and also very similar to those of the diffuse X-ray features in the NSC studied by \citet{2023MNRAS.522..635H}, which exhibit characteristic K-shell emission lines from Si, S, Ar, Ca and Fe, and an underlying continuum presumably due to bremsstrahlung of light elements. 
We restrict the spectral analysis over the energy range of 1.5--8 keV with XSPEC v12.14.1. 
We adopt the C-statistic for the spectral fit, only requiring a minimum of one counts per spectral bin.
For clarity of visualization, a heavier and adaptive binning is applied for the spectra presented in Fig.~\ref{fig:spectra}. 

\subsection{Spectral models}
\label{subsec:model}
As noted in \citet{2023MNRAS.522..635H}, the shock-heated stellar wind plasma may deviate from the collisional ionization equilibrium (CIE) state.
Hydrodynamical simulations also demonstrate that the temperature structures of CWB systems could be complex and the usual practice of a single-temperature model is likely invalid.
Our test of different spectral models demonstrated that an absorbed two-temperature, non-equilibrium ionization (NEI) model provides the best description of the present spectra, similar to the case of the NSC diffuse X-ray features studied in \citet{2023MNRAS.522..635H}.
The XSPEC model adopted can be expressed as {\it TBABS*DUSTSCAT*(VNEI+VNEI)}.
The Tuebingen-Boulder interstellar medium (ISM) absorption model ({\it TBABS} in XSPEC) is used to account for foreground absorption.
We also consider the foreground dust scattering (DUSTSCAT), by incorporating a multiplying factor of ${\rm exp}(-0.486 E^{-2}N_{\rm H})$, where $E$ is the photon energy in keV and $N_{\rm H}$ is the equivalent hydrogen column density in units of $10^{22}\rm~cm^{-2}$ \citep{1995A&A...293..889P}.
The temperatures and normalizations of the two $\it VNEI$ models are treated as free parameters during the spectral fitting. Also treated as free parameters are the abundances of Si, S, Ar, Ca and Fe.

As detailed in \citet{2023MNRAS.522..635H}, the composition of light elements, including H, He, C and N, determines the strength of the bremsstrahlung continuum, which in turn affects the measured heavy element abundances.
Following \citet{2023MNRAS.522..635H}, two scenarios of light element compositions are considered: a conventional (i.e., hydrogen-normal) scenario and a WR star wind (i.e., hydrogen-depleted) scenario.
The conventional scenario is commonly met in X-ray spectral analysis of the hot ISM, which assumes a solar abundance for all light elements especially hydrogen.
However, the WR stars, which just end their main-sequence phase, have evolved to synthesize a significant amount of carbon and nitrogen, while most of their hydrogen envelopes have been blown away. 
Depletion of hydrogen and enrichment of carbon and nitrogen have been widely observed in WR star winds \citep{2007ARA&A..45..177C,2014ARA&A..52..487S}.

Compared to the YNC, the Arches and Quintuplet clusters are younger and their WR stars are predominantly of the younger WN types.
In their winds, nitrogen abundances should be enhanced, while a larger fraction of hydrogen are preserved.
\citet{2008A&A...478..219M} measures the mass fraction of carbon and nitrogen in the Arches WR stars and provides the He/H ratios.
The average light element compositions of nearby infrared sources derived in \citet{2008A&A...478..219M} are used to represent the hydrogen-depleted scenario of the three Arches sources.
Since the WR stars in the Arches cluster have consistent spectral type and share similar He and N abundances, possible source confusion will not lead to significant changes in the light element compositions.
An estimation of the wind compositions of Edd 1 was also made by \citet{2009A&A...507.1567C}. 
So far, the wind properties of the Quintuplet WR stars lack useful measurements, thus we employ the average values of the Arches WR stars to represent Quintuplet, given their similarities.
Uncertainties introduced by this approximation is discussed in Section~\ref{subsubsec:Quintuplet}.
The two cases of light element compositions can be summarized as follows,

(i) {\it The hydrogen-depleted case}: $Z_{\rm H}/Z_{\rm He}/Z_{\rm C}/Z_{\rm N}$=$0.3/1/1/12$ (A1N), $0.5/1/1/9$ (A1S), $1/1/1/13$ (A2), $0.5/1/1/13$ (QX2), and $Z_{\rm H}/Z_{\rm He}/Z_{\rm C}/Z_{\rm N}$=$0.4/1/1/9$ (Edd 1).
Here as an example, $Z_{\rm H} = 0.3$ should be understood as a hydrogen number density 30 per cent of the reference standard of \citet{2000ApJ...542..914W}, and because hydrogen is depleted in a varied degree, He is taken as the reference element throughout.
The carbon abundance is assumed to be solar for all sources due to the absence of WC-type stars; nitrogen is significantly enriched in all sources. 
We note that A2 is barely hydrogen depleted. 

(ii) {\it The hydrogen-normal case}:
$Z_{\rm H}=Z_{\rm He}=Z_{\rm C}=Z_{\rm N}=1$ for all sources.

For each case, we proceed to the spectral fitting of the individual source spectra. 
The chosen two-temperature NEI model has 11 free parameters for each source, including the foreground column density, temperatures, ionization timescales and normalizations of the two temperature components and five heavy element abundances common between them. 
Given the relatively large amount of fitted parameters, we progressively free them in the fitting process to ensure that the best-fit model is achieved with a global (instead of local) minimum.
{\bf In our initial tests}, we find that the abundances in the three Arches sources are similar to each other, which is consistent with a common origin in the parent cluster. 
As a result, we tie the abundances across the three Arches sources in the formal spectral fit.

For Edd 1, the three high-state spectra are fitted simultaneously.
Upon comparison with the best-fit parameters of the Edd 1 low-state spectrum, we find similar temperatures and ionization timescales.
Also the abundances derived from the spectra in both states show little difference.
Considering the fact that the flux variation in Edd 1 (Figure \ref{fig:edd1_lc}) is likely caused by orbital modulated self-absorption, we further perform a joint fit to the spectra in both states, linking their abundances and leaving only the column density and the normalizations as independent parameters.

\begin{table*}
\centering
\renewcommand\arraystretch{1.3}
\setlength\tabcolsep{6pt}
\caption{X-ray Spectral Fit Results\label{tab:fitting_result}}
\begin{threeparttable}
\begin{tabular}{ccccc}
\hline
Model & Parameter & Arches (A1N/A1S/A2) & Quintuplet (QX2) & Edd 1 (high/low)  \\ 
\hline
H-depleted & $\rm{N_H}$ ($10^{22}\ \rm{cm}^{-2}$) & $10.66_{-0.57}^{+0.61}$ / $10.41_{-0.58}^{+0.62}$ / $10.15_{-0.68}^{+0.66}$ & $13.60_{-1.94}^{+2.07}$ & $11.19_{-0.63}^{+0.58}$ / $13.88_{-1.99}^{+2.05}$  \\ 
& Si & $1.21_{-0.41}^{+0.31}$ & $1.52_{-1.30}^{+1.17}$ & $1.38_{-0.48}^{+0.47}$  \\ 
& S & $0.75_{-0.17}^{+0.15}$ & $0.75_{-0.25}^{+0.24}$ & $1.38_{-0.21}^{+0.23}$  \\ 
& Ar & $0.60_{-0.21}^{+0.18}$ & $0.70_{-0.52}^{+0.50}$ & $1.39_{-0.20}^{+0.17}$  \\ 
& Ca & $1.35_{-0.25}^{+0.23}$ & $0.94_{-0.83}^{+0.83}$ & $1.37_{-0.20}^{+0.20}$  \\ 
& Fe & $1.63_{-0.21}^{+0.18}$ & $0.31_{-0.27}^{+0.25}$ & $1.47_{-0.13}^{+0.11}$  \\ 
& $\tau$ ($10^{11}\ \rm{cm}^{-3}\ \rm{s}$) & $8.78_{-7.03}^{+5.22}$ / $1.78_{-0.54}^{+0.66}$ / $3.55_{-1.12}^{+1.33}$ & $0.70_{-0.42}^{+0.53}$ & $3.74_{-0.77}^{+0.85}$  \\ 
& ${\rm kT_1}$ (keV) & $1.04_{-0.07}^{+0.13}$ / $1.54_{-0.21}^{+0.23}$ / $1.35_{-0.22}^{+0.19}$ & $0.89_{-0.24}^{+0.28}$ & $0.29_{-0.03}^{+0.03}$ / $0.66_{-0.25}^{+0.31}$  \\ 
& ${\rm Norm_1}$ ($10^{-3}\ \rm{cm}^{-5}$) & $1.07_{-0.22}^{+0.27}$ / $0.59_{-0.13}^{+0.13}$ / $1.55_{-0.14}^{+0.18}$ & $1.69_{-1.17}^{+1.39}$ & $3.10_{-1.03}^{+5.02}$ / $2.09_{-0.55}^{+1.91}$  \\ 
& ${\rm kT_2}$ (keV) & $5.55_{-1.48}^{+1.74}$ / $10.41_{-6.22}^{+13.77}$ / $5.22_{-1.24}^{+2.31}$ & $13.62_{-5.20}^{+14.23}$ & $1.95_{-0.13}^{+0.15}$ / $1.95_{-0.18}^{+0.19}$  \\ 
& ${\rm Norm_2}$ ($10^{-3}\ \rm{cm}^{-5}$) & $0.05_{-0.01}^{+0.02}$ / $0.02_{-0.01}^{+0.02}$ / $0.03_{-0.01}^{+0.04}$ & $0.05_{-0.04}^{+0.08}$ & $1.12_{-0.15}^{+0.13}$ / $0.74_{-0.22}^{+0.24}$  \\
& $L_{\rm 2-8~keV} (10^{33}\ \rm{erg}\ \rm{s}^{-1})$ & $1.32_{-0.01}^{+0.02}$ / $1.38_{-0.02}^{+0.02}$ / $2.96_{-0.04}^{+0.04}$ & $1.52_{-0.03}^{+0.02}$ & $3.58_{-0.05}^{+0.05}$ / $2.61_{-0.05}^{+0.06}$  \\ 
& $C/d.o.f. (\rm{P_{\rm Null}})$ & 1160.50/1133 (0.28) & 377.26/373 (0.43) & 1467.53/1451 (0.38)  \\ 
& $\chi^2/d.o.f.$ & 917.46/608 & 161.15/112 &  1271.89/936 \\

\hline
\hline
H-normal & $\rm{N_H}$ ($10^{22}\ \rm{cm}^{-2}$) & $10.18_{-0.64}^{+0.59}$ / $9.79_{-0.66}^{+0.58}$ / $10.19_{-0.63}^{+0.58}$ & $13.32_{-1.19}^{+1.47}$ & $10.89_{-0.61}^{+0.55}$ / $13.98_{-1.83}^{+1.97}$ \\
& Si  & $1.39_{-0.51}^{+0.43}$ & $1.77_{-1.48}^{+1.37}$ & $1.81_{-0.58}^{+0.62}$ \\
& S  & $1.05_{-0.19}^{+0.16}$ & $0.87_{-0.31}^{+0.26}$ & $1.88_{-0.27}^{+0.24}$ \\
& Ar & $0.80_{-0.23}^{+0.18}$ & $0.82_{-0.62}^{+0.55}$ & $1.95_{-0.33}^{+0.27}$  \\
& Ca & $1.46_{-0.28}^{+0.24}$ & $1.17_{-1.05}^{+0.78}$ & $1.96_{-0.31}^{+0.24}$  \\
& Fe & $1.70_{-0.20}^{+0.15}$ & $0.42_{-0.40}^{+0.38}$ & $2.08_{-0.18}^{+0.15}$  \\
& $\tau$ ($10^{11}\ \rm{cm}^{-3}\ \rm{s}$) & $10.06_{-5.14}^{+6.45}$ / $1.68_{-0.51}^{+0.74}$ / $3.36_{-0.68}^{+0.92}$ & $0.81_{-0.44}^{+0.43}$ & $3.41_{-0.73}^{+0.63}$  \\
& ${\rm kT_1}$ (keV)& $1.10_{-0.13}^{+0.17}$ / $1.75_{-0.26}^{+0.31}$ / $1.22_{-0.25}^{+0.28}$ & $0.84_{-0.35}^{+0.43}$ & $0.28_{-0.02}^{+0.03}$ / $0.54_{-0.15}^{+0.23}$  \\
& ${\rm Norm_1}$ ($10^{-3}\ \rm{cm}^{-5}$) & $0.68_{-0.18}^{+0.23}$ / $0.41_{-0.11}^{+0.13}$ / $0.96_{-0.11}^{+0.14}$ & $1.35_{-0.88}^{+1.09}$ & $4.06_{-0.78}^{+3.14}$ / $1.96_{-0.48}^{+2.51}$  \\
& ${\rm kT_2}$ (keV)& $4.87_{-1.77}^{+1.94}$ / $10.83_{-5.39}^{+11.63}$ / $5.06_{-1.52}^{+2.81}$ & $5.65_{-1.20}^{+1.43}$ & $2.03_{-0.15}^{+0.19}$ / $1.89_{-0.18}^{+0.19}$ \\
& ${\rm Norm_2}$ ($10^{-3}\ \rm{cm}^{-5}$) & $0.04_{-0.02}^{+0.02}$ / $0.04_{-0.02}^{+0.02}$ / $0.06_{-0.01}^{+0.02}$ & $0.07_{-0.02}^{+0.03}$ & $0.73_{-0.11}^{+0.17}$ / $0.51_{-0.10}^{+0.15}$ \\
& $L_{\rm 2-8~keV} (10^{33}\ \rm{erg}\ \rm{s}^{-1})$ & $1.30_{-0.01}^{+0.02}$ / $1.13_{-0.02}^{+0.02}$ / $2.85_{-0.04}^{+0.05}$ & $1.49_{-0.03}^{+0.02}$ & $3.77_{-0.06}^{+0.05}$ / $2.56_{-0.05}^{+0.06}$ \\
 & $C/d.o.f. (\rm{P_{\rm null}})$ & 1161.20/1133 (0.27) & 376.98/373 (0.43) & 1468.12/1451 (0.37) \\

& $\chi^2/d.o.f.$ & 916.39/608 & 162.03/112 &  1273.26/932\\

\hline
\hline
   & $N_{\rm photon}$  & 2386/2338/6072 & 2022 &  6516(I)+1073(G)+2425(S)/1178 \\
\hline
\end{tabular}
	\begin{tablenotes}
	\small
	\item
	Notes:
The upper and lower panels show the best-fit parameters with the hydrogen-depleted and hydrogen-normal cases, both using an absorbed two-temperature VNEI model. 

A joint fit is adopted for the three Arches sources as well as Edd1 in its two flux states. All free parameters are listed. 
The resulting Cash statistics are shown with the corresponding null hypothesis probablities and the $\chi^2$-statistic results are shown for comparison.
The last row gives the number of photons in the fitted spectra. The quoted errors are at the 90\% confidence level.
See text for details on the spectral models.
	\end{tablenotes} 

\end{threeparttable}
\end{table*}

\begin{figure*}
    \centering
    \includegraphics[width= 0.48\textwidth]{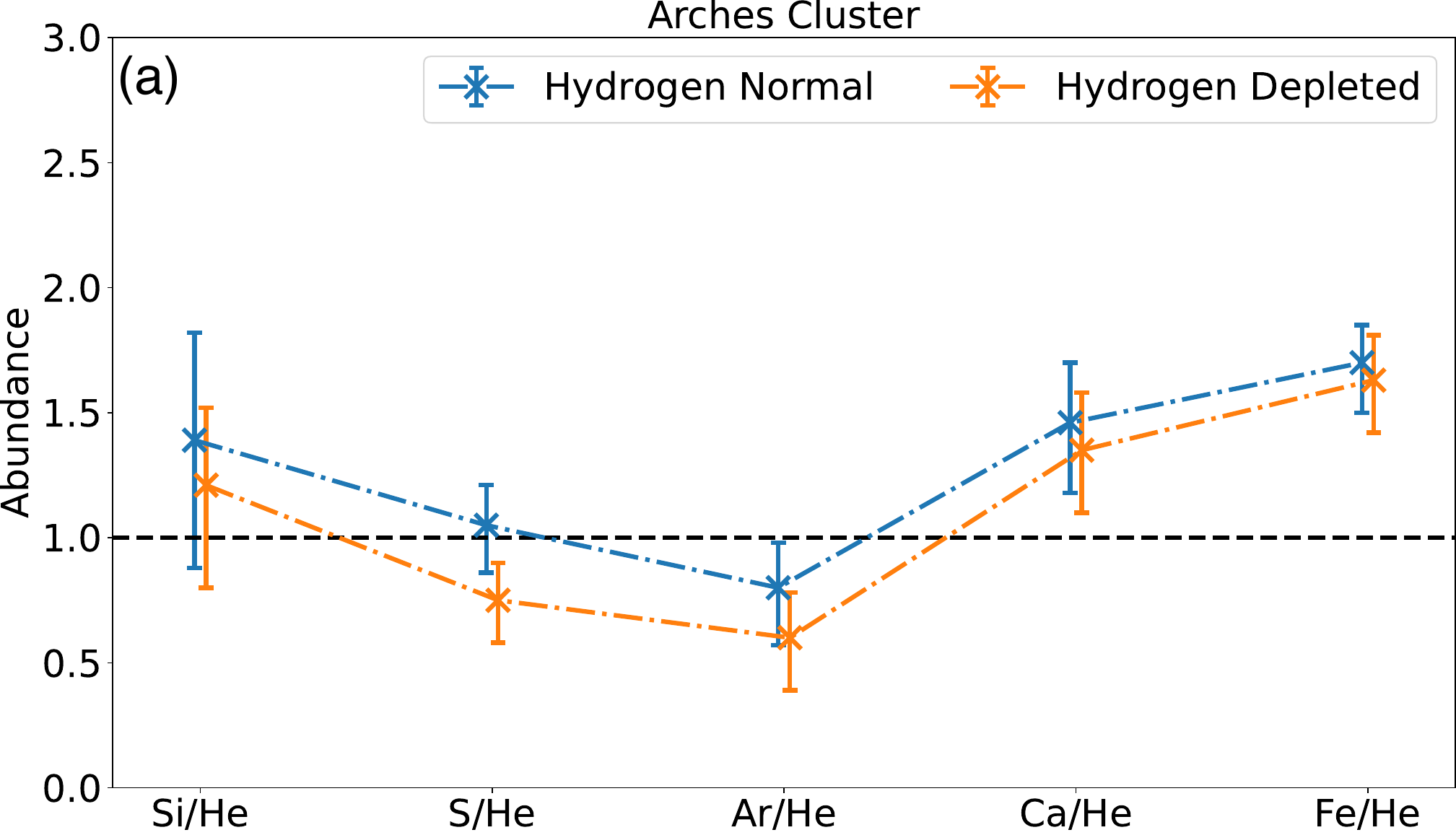}
    \includegraphics[width= 0.48\textwidth]{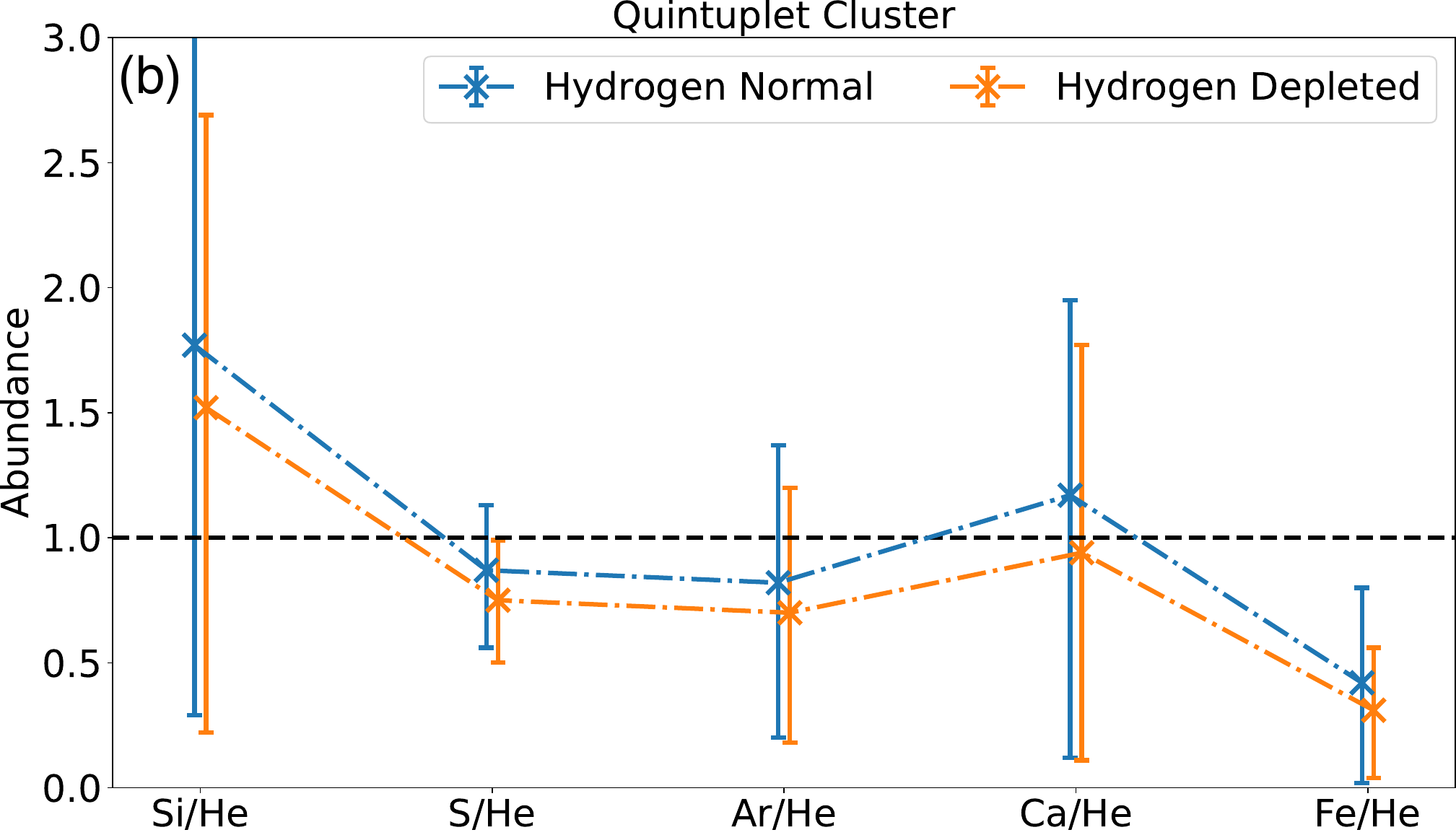}
    \includegraphics[width= 0.48\textwidth]{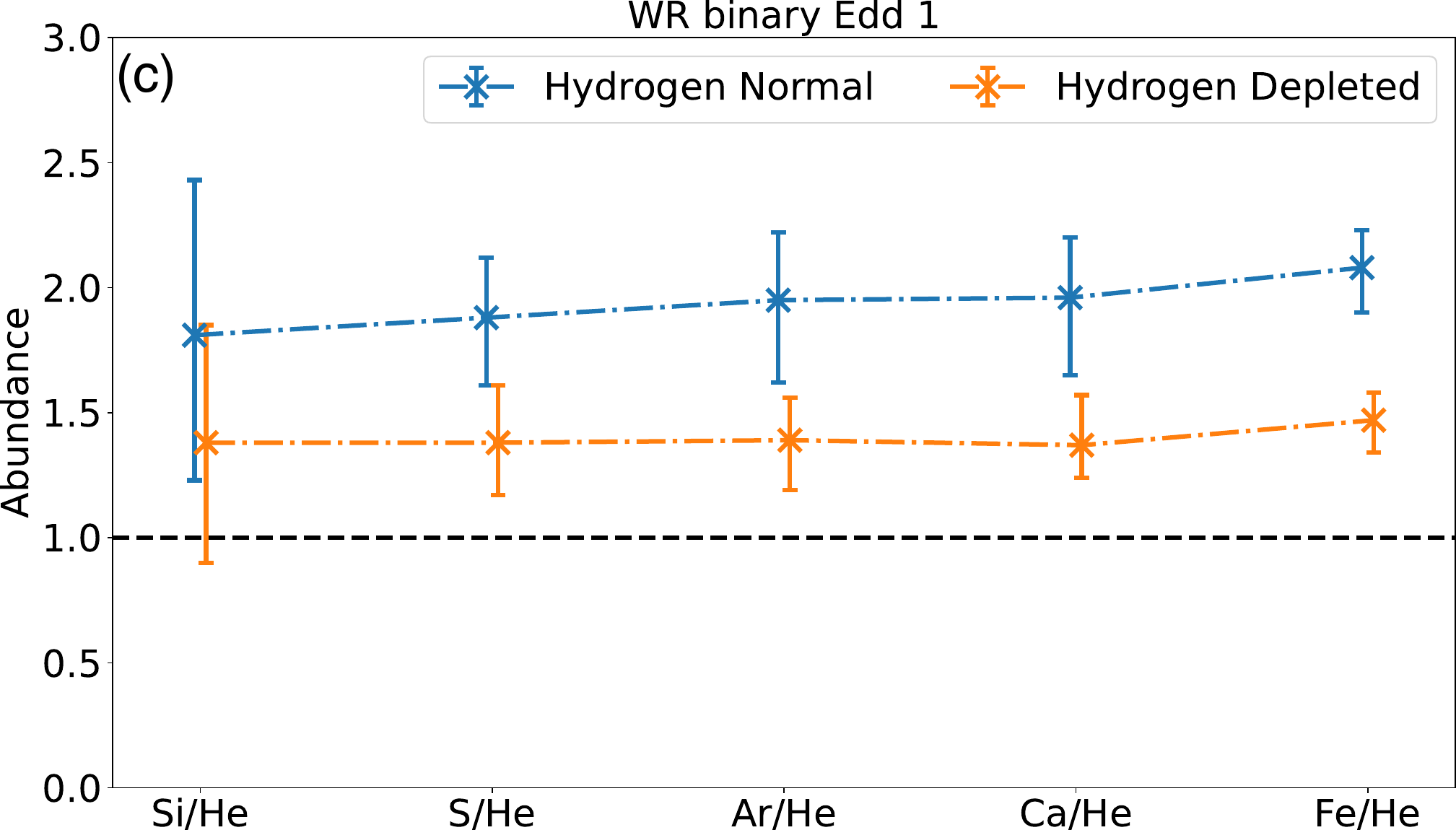}
    \includegraphics[width= 0.48\textwidth]{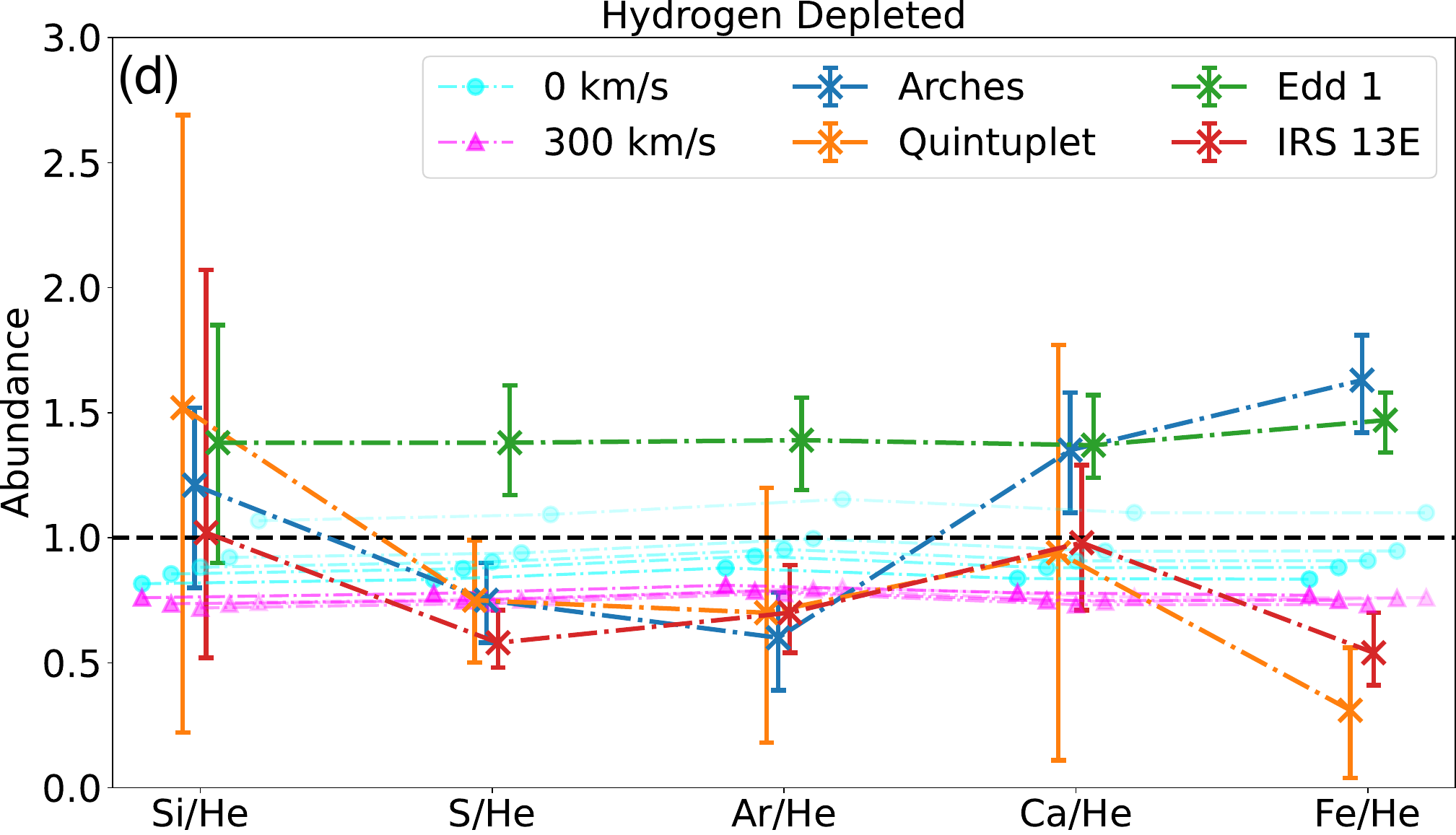}
    \caption{The best-fitted abundances of the heavy elements. Panels (a), (b) and (c) show the measured abundances in Arches, Quintuplet and Edd 1, respectively. 
    The hydrogen-depleted and hydrogen-normal case is marked with orange and blue, respectively.
    In panel (d), the abundances in Arches, Quintuplet and Edd 1 are derived as described in text and labeled with blue, orange and green, respectively. The hydrogen-depleted case is assumed.
    The error bars are at the 90\% level.
    The measured abundances of IRS 13E, adopted from \citet{2023MNRAS.522..635H} as a YNC representative, are shown in red. Different sets of abundances are slightly shifted along the x-axis for better clarity.
    The model-predicted heavy element abundances of the WR star winds from \citet{2024ApJS..272...15R} are also plotted. These WR stars are assumed to have a 2 solar metallicity, an equatorial rotation velocity of 0 km/s (cyan curves, i.e., non-rotation) or 300 km/s (magenta curves, i.e., fast rotation), and an initial mass ranging between 30 -- 120 $\rm M_\odot$ (lighter to darker color).
    The horizontal black dashed line marks the solar value.}
    \label{fig:abundance}
\end{figure*}

\begin{figure*}
	\includegraphics[width= 0.48\textwidth]{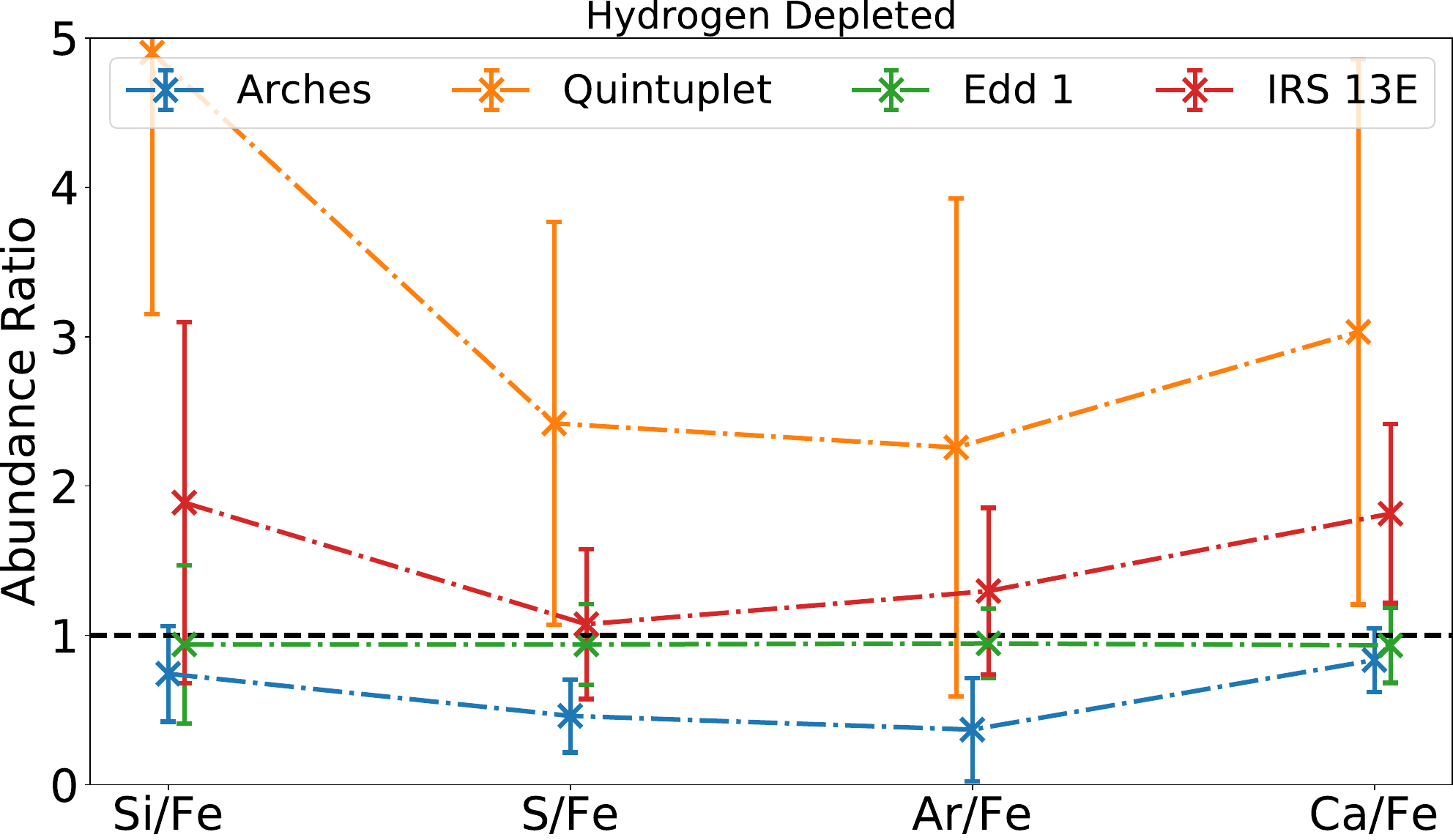}
    \includegraphics[width= 0.48\textwidth]{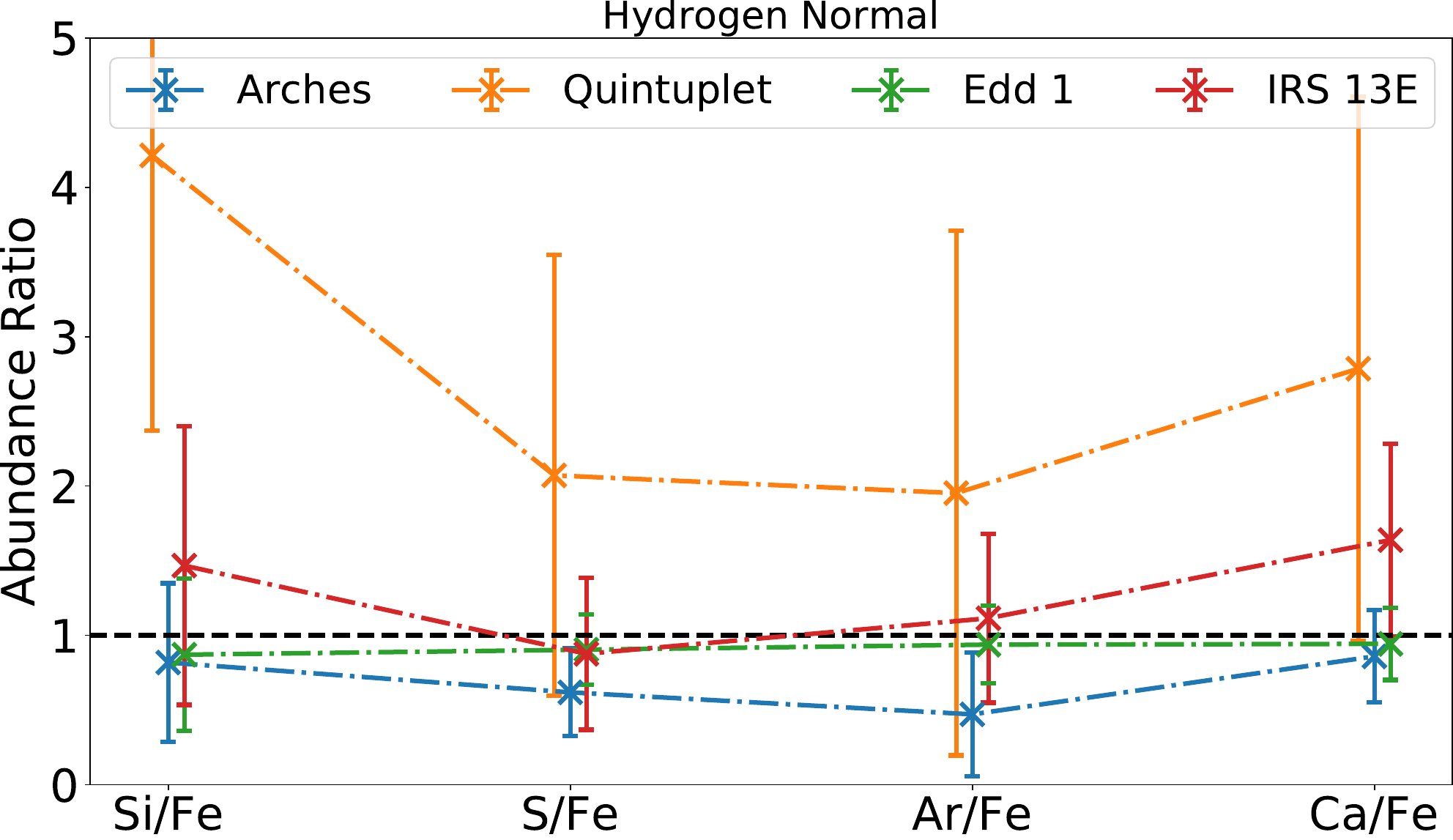}
    \caption{The best-fitted $\alpha$/Fe abundance ratio. The left and right panels show the hydrogen-depleted and hydrogen-normal cases, respectively. The different color symbols represent different sources, which are slightly shifted along the x-axis for clarity. The error bars are at the 90\% level. The horizontal black dashed line marks the solar value.}
    \label{fig:alpha_ratio}
\end{figure*}

\subsection{Fitting results}
\label{subsec:results}
Based on the best-fit two-temperature NEI model, the heavy element abundances are derived and summarized in Table~\ref{tab:fitting_result}.
Estimations of the unabsorbed 2--8 keV luminosity based on the best-fit parameters are also provided in Table~\ref{tab:fitting_result}.
The best-fit foreground column density is $N_{\rm H}\sim 10\times10^{22}\rm~cm^{-2}$ in Arches and $N_{\rm H}\sim13\times10^{22}\rm~cm^{-2}$ in Quintuplet.
A clear difference in column density can be found between the high- and low-state spectra of Edd 1, supporting an orbital motion-induced variation.
In both Arches and Quintuplet, the best-fit temperature is $\sim 1~{\rm keV}$ for the low-temperature component and $\sim5~{\rm keV}$ for the high-temperature component.
A lower temperature is found in Edd 1, with values of $0.3 - 0.7~{\rm keV}$ and $\sim 2~{\rm keV}$ for the two temperature components.
The ionization timescales are on the order of $10^{11}{\rm~cm^2~s}$, lower than the characteristic ionization equilibrium timescales \citep{2010ApJ...718..583S}.
A larger ionization timescale is found in A1N, though it still does not reach the equilibrium state.

Elemental abundances derived based on the hydrogen-normal and hydrogen-depleted cases are shown in Fig.~\ref{fig:abundance}(a)-(c) for Arches, Quintuplet and Edd 1, respectively.
Despite the significant statistical uncertainties in the individual elements, it is evident that the hydrogen-depleted case shows a systematically low abundance for all five elements, compared to the hydrogen normal case.
This can be understood as due to a higher contribution of helium (thus a relatively lower abundance of the heavy elements with respect to He) to compensate for the loss of hydrogen in contributing to the bremsstrahlung continuum.
The Arches sources show generally super-solar abundances of Si, Ca and Fe as well as a near-solar abundance of S and Ar.
In contrast, the Quintuplet source exhibits a sub-solar Fe abundance, while the other elements are near-solar.
In Edd 1, all five elements have a strong and rather uniform enrichment, with an abundance of $\sim2$ solar.

For a more direct comparison, we plot the Si, S, Ar, Ca and Fe abundances of the difference sources in Fig.~\ref{fig:abundance}(d), adopting the hydrogen-depleted case.
The NSC/YNC source IRS 13E, the X-ray emission of which is thought to be originated from the colliding wind of a pair of WR stars \citep{2020ApJ...897..135Z}, is also included in the plot as a YNC representative.
It is worth noting that the abundances measured for the other three diffuse X-ray features in the NSC are broadly consistent with IRS 13E \citep{2023MNRAS.522..635H}.
It is evident that the three young massive clusters in the GC share a similar pattern of the $\alpha$ elements, with a sub-solar abundance of S and Ar, and a near- or moderately super-solar abundance of Si and Ca, although with substantial uncertainties.
On the other hand, their Fe abundance varies significantly, showing both subsolar (YNC and Quintuplet) and supersolar (Arches) values. 
Finally, Edd1 shows a markedly different abundance pattern, which suggests a different origin for this isolated WR binary, perhaps related to its non-cluster nature.

In addition to the absolute abundances, the abundance ratio between the $\alpha$ elements and iron, $\alpha$/Fe, is also an important indicator of the star formation properties and metal enrichment history \citep{1997ARA&A..35..503M,2013ARA&A..51..457N}.
We calculate $\alpha$/Fe and its associated uncertainties using the bootstrapping method, employing the XSPEC function ${\it fakeit}$.
The results of this calculation for both the hydrogen-depleted and hydrogen-normal cases are shown in Fig.~\ref{fig:alpha_ratio}.
Because the abundance ratios are insensitive to the underlying continuum, the two cases yield similar results, as evident in Fig.~\ref{fig:alpha_ratio}.
Interestingly, the four sources together exhibit a decreasing trend in $\alpha$/Fe, with Quintuplet being the most supersolar ($\sim3$), followed by IRS 13E (also supersolar), Edd 1 (near-solar) and Arches (subsolar) in order. 
We suspect that the elevated $\alpha$/Fe ratios in Quintuplet may result from dust depletion of Fe, a possibility further addressed in Section~\ref{subsubsec:Quintuplet}.

\section{Discussion}
\label{sec:Discussion}

\subsection{Caveat in the abundance measurements}
\subsubsection{Potential contamination to the observed spectra}
Other objects and radiation mechanisms, including accreting compact objects, thermal emission from main-sequence stars, reflection from the molecular gas, and non-thermal emission from synchrotron and/or inverse Compton scattering can also contribute to the observed X-ray emission, in particular the continuum, which in turn affects the determination of the relative strength of the metal lines. 

Arches and Quintuplet are probably too young for their member stars to evolve to the compact object (i.e., black hole or neutron star) stage, and so far there is no proven evidence for compact objects in either cluster.
Also in these two clusters, the WR stars concentrate in the core region, and their X-ray luminosity plays a dominant role over that of main-sequence stars. 
The thermal X-ray emission from main-sequence stars has a relatively low temperature and should contribute little to the 2--8 keV band.

\citet{2006MNRAS.371...38W} investigated the global X-ray spectral properties of the Arches cluster and found prominent 6.4 keV Fe fluorescence lines.
They suggested that the 6.4 keV line arises from the interaction between low-energy cosmic ray electrons and the ambient medium.
However, in our background-subtracted spectra, the existence of the 6.4 keV line lacks evidence.
This indicates that the reflected emission from cold gas is uniform in the source and background extraction regions and is effectively removed through background subtraction.

In the colliding wind zone, relativistic electrons can be generated and radiate through synchrotron and inverse Compton scattering.
These non-thermal emission from CWB systems has been observed in the radio, hard X-ray and $\gamma$-ray bands \citep{2013A&A...558A..28D,2018NatAs...2..731H,2009ApJ...698L.142T,2020A&A...635A.167H}.
However, hard X-ray spectral fitting of known CWBs, including Apep \citep{2023A&A...672A.109D}, $\eta$ Car \citep{2018NatAs...2..731H} and $\gamma$ Vel \citep{1995A&A...298..549W}, indicates that the non-thermal component has little effect on the 0.2--10 keV band.
\citet{2024arXiv241017806C} observed the Arches cluster with VLA at 6 and 10 GHz in several epochs. 
The radio sources (F6, F7 and F9) detected closest to the three X-ray sources share a flat-to-negative spectral index, indicating non-thermal radiation. 
Their strong flux and spectral index variations reinforce the colliding wind origin.
The average flux densities of these radio sources are on the order of $0.1\text{-}1~\rm{mJy}$.
Assuming the same electron population, magnetic field and radiation field as those assumed for the NSC by \citet{2023MNRAS.522..635H}, we find that the non-thermal 2--8 keV flux extrapolated from the radio frequencies is 2--3 orders of magnitude lower than the observed values, which means that the non-thermal component is negligible for the Arches sources. While less is known about the radio properties of QX2 and Edd 1, any non-thermal emission is similarly unlikely to have a significant contribution to their observed X-ray spectra. 

\subsubsection{Potential bias in Quintuplet}
\label{subsubsec:Quintuplet}
When fitting the Quintuplet spectrum, an assumption is made that uses the same light element compositions as Arches in the hydrogen-depleted case, since there is no precise near-infrared measurement of the light element compositions for the Quintuplet WR stars.
This implicitly assumes that the WR star population in the Arches and Quintuplet clusters are at nearly the same evolutionary stage. 
In fact, Quintuplet is slightly older and hosts a more evolved WR star population, with a higher fraction of WC-type stars \citep{2018A&A...618A...2C}.
With regard to QX2, the closest near-infrared counterpart is a B0I star (source No.257 in \citealp{1999ApJ...514..202F}) according to \citet{2006MNRAS.371...38W}.
As shown in Fig.~\ref{fig:xray_image}(c), there are no known WR stars located exactly at the location of QX2.
The three nearest WR stars have spectral types of WC9d, WN9/WN11h and WC9d, respectively \citep{2009A&A...494.1137L}.
The more evolved WC-type stars will generate winds containing more carbon and less hydrogen. 
Considering this effect, there should be less hydrogen and nitrogen, as well as more carbon contributing to the total bremsstrahlung continuum. 
However, the small number ratio of carbon to hydrogen means that the extra carbon cannot compensate for the reduction of hydrogen, which would imply for an even lower heavy element abundance. 
For instance, reducing the assumed hydrogen and nitrogen content by half and increasing the carbon content by a factor of two will lead to a $\sim$25\% lower abundance of the heavy elements.
Thus the qualitative behavior of QX2 discussed in Section~\ref{subsec:results} remains valid. 

The WC-type star is empirically more likely to produce dust in their winds \citep{2024arXiv241004436S}.
Two of the three nearest WR stars of QX2 have a spectral type of WC9d, where ``d'' refers to persistent dust formation \citep{2001AJ....121.2115S}.
Therefore dust depletion might have a non-negligible impact on the element abundances.
Cosmic dust is primarily composed of olivine and pyroxene \citep{2007A&A...462..667M,2016ApJ...830...71F}, with O, Si, Ca, Fe as their main components.
When these elements are deposited into dust, their gas-phase abundances would be significantly reduced.
As mentioned in Section~\ref{subsec:results}, QX2 is unique in terms of a low Fe abundance and a high $\alpha$/Fe ratio, which can be understood as a result of dust depletion of Fe.
Significant dust is also present in the colliding wind zone of IRS 13E \citep{2010ApJ...721..395F,2020ApJ...897..135Z}, which has the second largest $\alpha$/Fe after QX2.
Based on the \citet{2000ApJ...542..914W} abundance standard and the best-fit spectral model normalization of QX2, we estimate a Ca plus Fe mass of $\sim10^{-6}~\rm{M_{\odot}}$, which is consistent with inferred dust mass of CWBs based on infrared observations \citep{2020ApJ...898...74L}.
However, although Si is also a major composition of the dust, we do not find a low abundance of Si in QX2, compared to the other WR stars analyzed.

\subsection{Inferring the initial stellar metallicity}
\label{sec:Difference}
At face value, the subsolar or near-solar abundances of the heavy elements derived for the three young star clusters (Fig.~ \ref{fig:abundance}) are rather surprising. 
If one considers that star formation has continuously or episodically taken place in the Galactic center, at least in the recent past, one would expect  substantial metal enrichment to the CMZ, the natural fuel of new stars.
However, the derived heavy element abundances, which essentially reflect the metallicity of the WR star winds, 
are not necessarily the same as the initial metallicity of the progenitor massive star when it just begins the main-sequence stage.

In principle, a link between the wind metallicity and the initial stellar metallicity can be quantitatively established, provided detailed modeling of stellar evolution and nucleosynthesis as well as modeling of the wind formation at the WR stage.  
This was recently offered by \citet{2018ApJS..237...13L}, who used a set of pre-supernova massive star models with masses between 13 and 120 ${\rm M_{\odot}}$ to study their evolutionary history and elemental yields.
The yields of these stars, reflecting the chemical compositions of their stellar winds, are presented assuming solar and sub-solar initial metallicities.
A follow-up work of \citet{2024ApJS..272...15R} extends this study to include modelled stars with an initial metallicity of two times solar.
Based on the yields of massive stars with an initial mass of 30--120 ${\rm M_{\odot}}$ (a practical mass range for the formation of WR stars) predicted by the supersolar models of \citet{2024ApJS..272...15R}, we calculate the abundances of Si, S, Ar, Ca and Fe (with respect to He) and plot them in Fig.~\ref{fig:abundance}(d) for comparison. 
Two sets of model predictions are included, which correspond to an equatorial rotation velocity of $0~{\rm km~s^{-1}}$ (i.e. non-rotating) and $300~{\rm km~s^{-1}}$ (i.e. fast-rotating). Within each set, a higher initial mass results in a smaller wind abundance. 
On the other hand, the predicted wind abundances of the non-rotating stars are systematically higher than the fast-rotating ones.
According to \citet{2018ApJS..237...13L}, stars with higher initial masses exhibit more confined convective zones, preventing the full mixing of heavy elements. 
Rotation further increases the molecular weight in the radiative envelope and decreases the opacity in the mantle, leading to less extended convective cores. 
Remarkably, these models show that the abundances of the five heavy elements in the WR star winds can be reduced to about half of their initial values, suggesting that 
a massive star with a solar or supersolar initial metallicity can still produce a subsolar metallicity wind after taking into account the nucleosynthesis and internal mixing of the star.
The predicted abundances are broadly consistent with the measured values for the Arches, Quintuplet and IRS 13E, as evident in Fig.~\ref{fig:abundance}(d).
In other words, the initial metallicity of the three star clusters could well be supersolar, compatible with the physical expectation that the GC region is generally metal-enriched.
On the other hand, this effect will have little influence on the $\alpha$/Fe abundance ratio since the depletions of different heavy elements in the stellar wind are similar to each other. 

As for Edd 1, even if we take into account the difference between the wind metallicity and initial metallicity, its measured heavy element abundances ($\sim1.5$ solar) is still significantly higher than the model predictions.
By face value, this could imply an initial metallicity of $\gtrsim 3$ solar.
It is unclear whether such a high metallicity is easily achievable in the CMZ, or perhaps Edd 1 was born in a molecular cloud of an exceptionally high metallicity.
Future abundance measurements of more isolated massive stars in the GC would help understand this issue.

\subsection{Implications on the Galactic center environment}
\subsubsection{The young massive clusters}

It is found that the WR stars in the three young clusters have similar Si, S and Ar abundances but different Ca and Fe abundances. 
In particular, while Arches has a supersolar Fe abundance, Quintuplet (QX2) is more consistent with the YNC (IRS 13E), both having a subsolar Fe abundance.
As mentioned in Section~\ref{subsubsec:Quintuplet}, the dust depletion effect can be responsible for this discrepancy.
The similar abundances found in the three clusters indicate a common gas reservoir for their formation.
For Arches and Quintuplet, which are part of the NSD, they are naturally born from the abundant molecular gas in the CMZ. 
The original fuel for the YNC, on the other hand, is less certain, as little molecular gas is currently found connected to the YNC \citep{2020A&A...641A.102S,2023ASPC..534...83H}.
The similar abundances would favor an origin of an infallen molecular cloud into the central parsecs, part of which later become new stars despite the strong tidal force of Sgr~A*. But this is certainly an oversimplified picture.  For instance, stellar feedback could well produce a local effect in altering the metallicity of neighboring molecular clouds, with little influence on a larger scale.

\subsubsection{Edd 1 and isolated WR binaries}
The isolated WR binary Edd1 locates 4 arcmin away from the central SMBH, Sgr A*,  which corresponds to a projected physical distance of $\sim$10 pc.
The markedly different abundances found in Edd~1 provides a hint on the origin of the isolated WR stars distributed across the GC.
Previous studies claim that some of these isolated WR stars could have been former members of Arches and later kicked out due to dynamic interactions \citep{2014A&A...566A...6H,2023MNRAS.521.4473C}.
Other works suggest that these isolated WR stars used to belong to another cluster that has been tidally dissolved by the strong tidal shearing in the GC \citep{2020MNRAS.495.1209R,2022NatAs...6.1178N,2024A&A...683A...3M}.
Also part of these isolated WR stars could have formed in isolation in the dense molecular clouds.
The discrepancy we find between the abundances in Edd 1 and Arches disfavors the Arches member hypothesis.
Whether Edd 1 is originated from a previously tidally-dissolved cluster or formed in isolation, its significantly higher abundances suggest a distinct origin compared to the three currently known young massive clusters.

\subsubsection{The accretion flow onto Sgr A*}
\citet{2013Sci...341..981W} analyzed the quiescent X-ray spectrum of Sgr A*, taken from the {\it Chandra} ACIS-S/HETG zeroth-order non-disperse data, and found that a radiatively inefficient accretion flow model with a 1.5 solar abundance can provide a good fit to the observed spectrum.
\citet{2020ApJ...891...71C,2024ApJ...974...98B,2024ApJ...974...99B} further investigated the HETG first-order disperse spectrum to constrain the accretion flow properties.
\citet{2024ApJ...974...98B} found that the X-ray data requires a sub-solar Fe abundance of $Z_{\rm Fe}<0.32$.
Extensive hydrodynamic simulations have demonstrated that Sgr A* is currently fed by a circumnuclear hot gas, which originates from the colliding winds of the WR stars within the YNC \citep[e.g.][]{2018MNRAS.478.3544R,2020MNRAS.493..447C,2020ApJ...888L...2C}.
Therefore, the accretion flow onto Sgr~A* should inherit the elemental abundances of the WR star winds, which, however, were not considered in the aforementioned works.
In particular, one expects that the accretion flow is hydrogen depleted to some extent, simultaneously with an enrichment of carbon and nitrogen and a reduction of heavy elements. This should have a significant effect on the inferred properties of the accretion flow.

As a simple demonstration, we fit the HETG non-disperse spectrum of Sgr A* in its quiescent state to assess the differences.
As shown in Fig.~\ref{fig:sgra_spec}, the spectra are extracted following the same procedure described in \citet{2013Sci...341..981W}.
Due to the similarities between the Sgr A* spectrum and the WR star spectra investigated in the above, the same two-temperature NEI model is adopted in the spectral fit.
As described in Section~\ref{subsec:model}, a total number of 11 free parameters are included.
The hydrogen-normal and hydrogen-depleted cases are examined, and both yield a good fit to the observed spectrum.
The NEI model normalizations, $\frac{10^{-14}}{4\pi[D_{A}(1+z)]^2}\int n_{\rm e} n_{\rm H}{\rm dV}$, can provide an estimate of the average gas number density of the accretion flow.
The relation between $n_e$ and $n_H$ is calculated on the basis of the light element compositions, and
a spherical volume with a radius of 0.06 pc (corresponding to the spectral extraction region and roughly the Bondi radius, \citealp{2013Sci...341..981W}) is assumed.
It is further assumed that the two temperature components each occupy  a fraction of the volume under pressure equilibrium (i.e. constant $n\times T$).
The resultant mean gas density (hence the overall mass accretion rate) of the accretion flow is quite different between the two cases: $1.0~{\rm M_{\odot}~ pc^{-3}}$ for hydrogen-depleted and $1.6~{\rm M_{\odot}~pc^{-3}}$ for hydrogen-normal.
This strongly suggests that the peculiar elemental abundances inherited from the WR stars should be taken into account in future studies of the accretion flow onto Sgr~A*.
\begin{figure}
    \includegraphics[width= 0.48\textwidth]{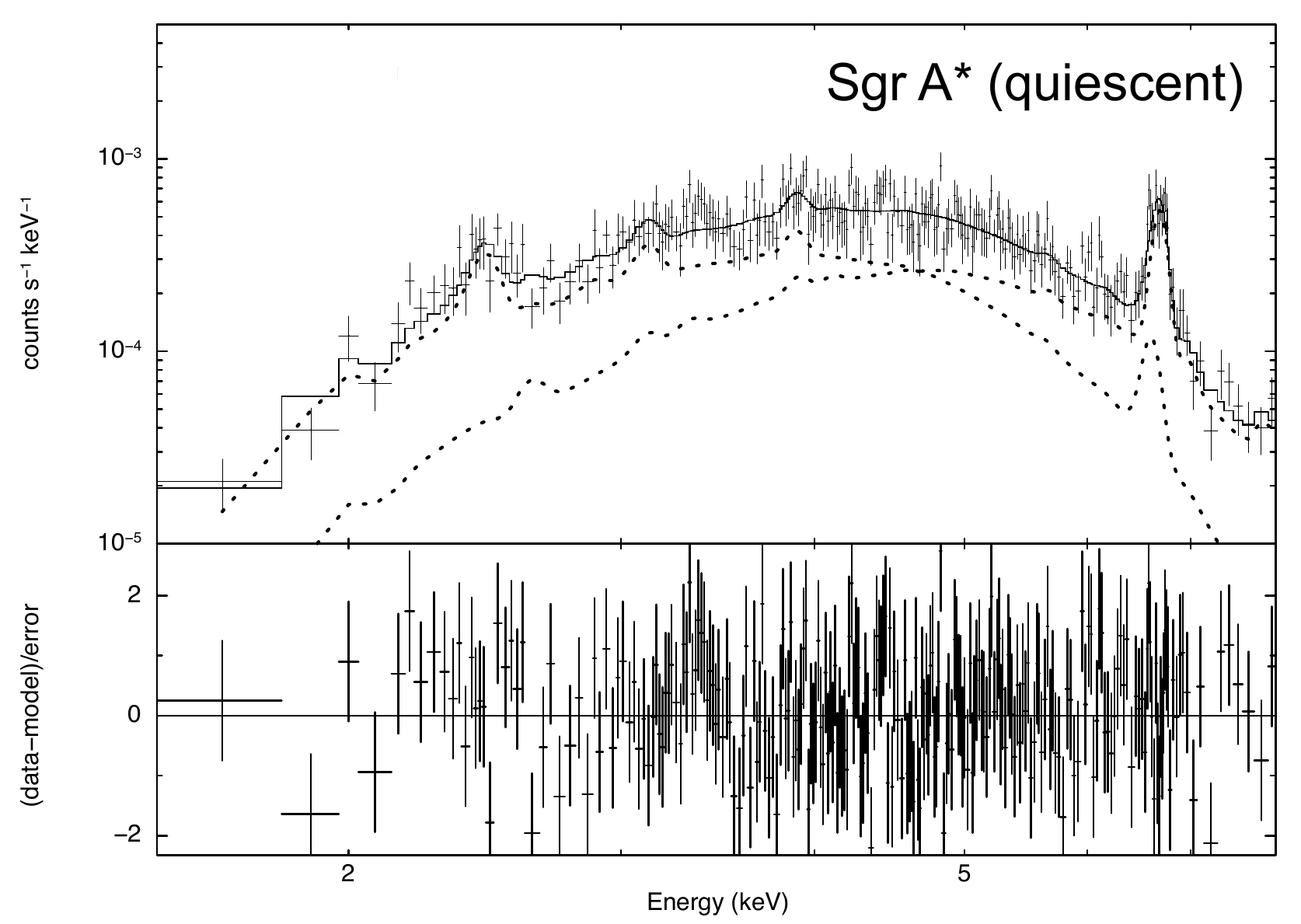}
    \caption{The background-subtracted ACIS-G spectrum of the SMBH Sgr A*. An adaptive binning has been applied to achieve at least 10 counts and S/N greater than 3 in each spectral bin. The best-fit XSPEC model, ${\it TBABS*DUSTSCAT*(VNEI+VNEI)}$, is shown by the solid line, with the two NEI components shown by the dotted lines, assuming a hydrogen-depleted case. The bottom panels show the relative residuals. The error bars are at the 1$\sigma$ level.}
    \label{fig:sgra_spec}
\end{figure}

\section{Summary}
\label{sec:sum}
In this work, based on deep {\it Chandra} observations, we have performed the first X-ray measurement of heavy element (Si, S, Ar, Ca and Fe) abundances of a few WR stars in the Galactic center, which reside in massive star clusters or in the field. The measurements are derived from a good fit to the observed X-ray spectra using a two-temperature NEI model, taking into account the fact that WR star winds are depleted in hydrogen, which is consistent with our previous analysis for the diffuse X-ray features in the NSC, thus allowing for a direct comparison. Our main findings include:

\begin{itemize}
    \item The WR stars in Arches and Quintuplet show similar  abundances of Si, S and Ar, which are also similar to those of IRS 13E, a colliding wind system representative of the young nuclear cluster. Quintuplet and IRS 13E show a low abundance of Ca and Fe, which could be due to significant dust depletion at a more evolved WR phase. These similarities suggest a common gas reservoir for the formation of the three clusters. 
    
    \item The field source Edd~1 has a significantly higher and semi-uniform abundance for all five elements, reaching $\sim$1.5 solar, disfavoring previous suggestion that it was a former member of Arches and implying for a different origin. 

    \item The sub- to near-solar abundances of the WR star winds can be naturally understood as the result of nucleosynthesis and internal mixing of the parent star, which have a supersolar initial metallicity as expected for the GC region. 
\end{itemize}

\section*{Acknowledgements}
This work is supported by the National Key Research and Development Program of China (No. 2022YFF0503402), the National Natural Science Foundation of China (grant 12225302), and the CNSA program D050102.
The authors wish to thank Zhao Su, Fangzheng Shi and Zhenlin Zhu for helpful discussions.

\section*{Data Availability}
The data underlying this article will be shared on reasonable request to the corresponding author.
\bibliographystyle{mnras}
\bibliography{main.bib}

\appendix

\section{{\it Chandra} Observation Log}

Table~\ref{tab:obs_info} lists the {\it Chandra} observations adopted to obtain the spectra of Arches and Quintuplet. In the ''Note'' column, "A" denotes the observations used for the Arches cluster, and "Q" denotes those used for the Quintuplet cluster. The ''Exposure'' column indicates the actual exposure time and does not account for filtered time intervals due to background flares.

\section{The X-ray spectrum of QX6}\label{QX6}
Fig.~\ref{fig:qx6} shows the background-subtracted spectrum of QX6, which exhibits significant emission lines between 6--7 keV. 
We add an absorbed bremsstrahlung continuum plus two Gaussian lines (${\it TBABS*PCFABS*(BREM+GAUSS+GAUSS)}$ in XSPEC) to the spectrum, fixing the Gaussian centroids at 6.7 and 7.0 keV in accordance with Fe XXV K$\alpha$ and Fe XXVI Ly$\alpha$ lines.
The fitted bremsstrahlung temperature is $3.4^{+2.8}_{-0.9}~{\rm keV}$, while the unabsorbed 2--8 keV luminosity based on the best-fit model is $8.7_{-3.0}^{+12.5}\times10^{31} {\rm~erg~s^{-1}}$ (assuming a distance of 8 kpc).
These X-ray spectral properties are consistent with cataclysmic variables, which are found to be prevalent in the GC (e.g. \citealp{2009ApJS..181..110M}).
Given the youth of the Quintuplet cluster (3-4 Myr), QX6 is unlikely to be a genuine member assuming its CV nature. 
Far more plausible is a chance alignment along our line of sight.
This interpretation is supported by its markedly lower foreground column density,$N_{\rm H}\approx6\times10^{22}~{\rm cm^{-2}}$, compared with bona‑fide cluster sources.

\begin{figure}
    \includegraphics[width= 0.48\textwidth]{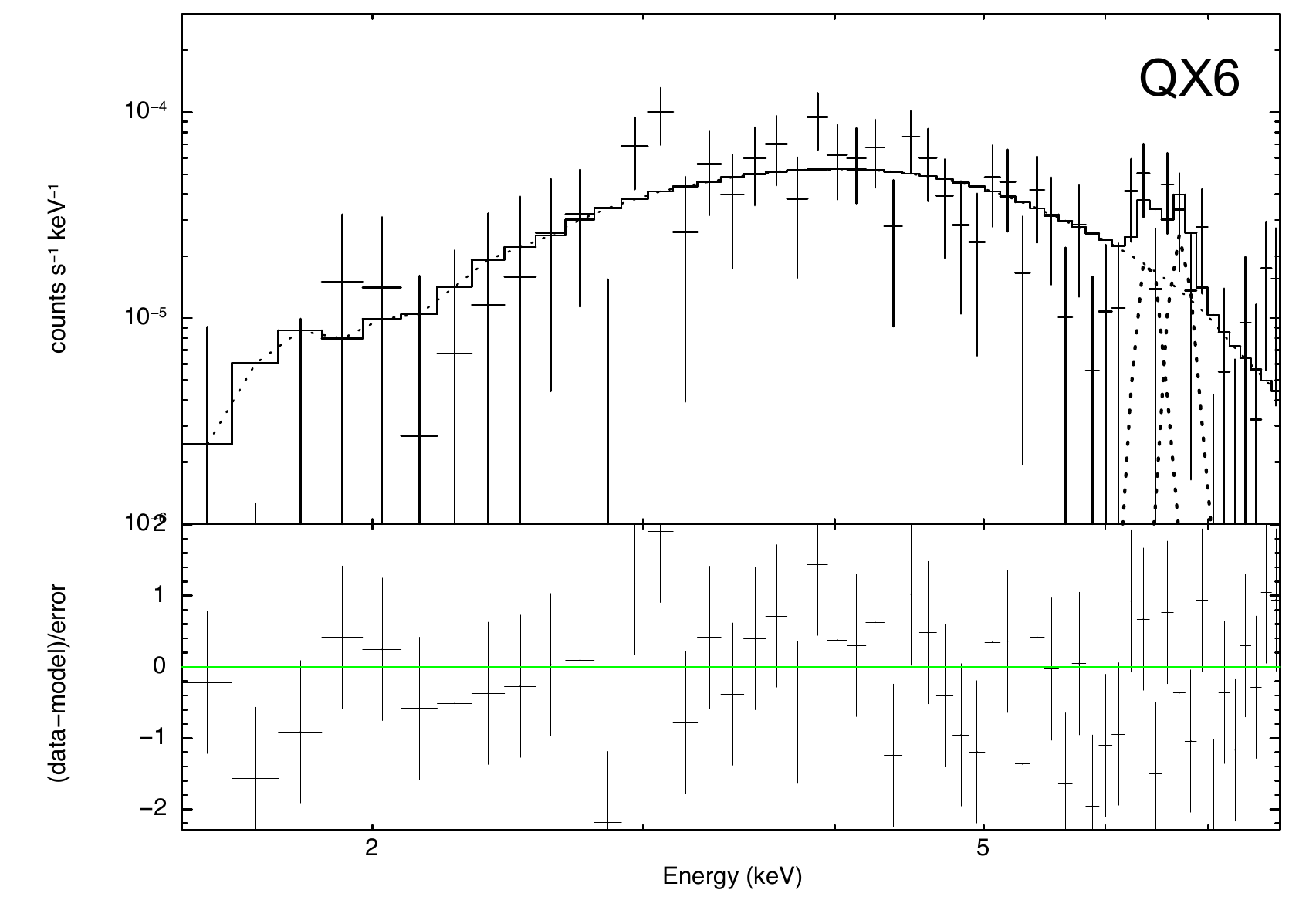}
    \caption{The background-subtracted spectrum of QX6 in the Quintuplet cluster. An adaptive binning has been applied to achieve at least 10 counts and S/N greater than 3 in each spectral bin. The best-fit XSPEC model, ${\it TBABS*PCFABS*(BREM+GAUSS+GAUSS)}$, is shown by the solid line. The bottom panels show the relative residuals. The error bars are at the 1$\sigma$ level.}
    \label{fig:qx6}
\end{figure}

{\small
\onecolumn
\begin{longtable}{@{\extracolsep{\fill}}lcccccc@{}}
\caption{{\it Chandra} Observations of Arches and Quintuplet} \label{tab:obs_info} \\

\hline
Observation ID & Start Time (UT) & Exposure (ks) & Target R.A. (J2000) & Target Decl. (J2000) & Roll Angle (deg) & Note\\
\hline
\endhead

\hline
\endfoot

\hline
\endlastfoot
945 & 2000-07-07    19:05:19 & 48.8 & 266.5809 & -28.8717 & 284.4    & A,Q\\
2273 & 2001-07-18    00:48:28 & 11.6 & 266.7082 & -28.8757 & 283.8   & Q\\
2276 & 2001-07-18    04:16:58 & 11.6 & 266.5180 & -28.7743 & 283.8   & Q\\
4500 & 2004-06-09    08:50:32 & 98.6 & 266.4838 & -28.8195 & 55.2    & A,Q\\
6311 & 2005-07-01    02:05:26 & 4.0 & 266.3680 & -28.8546 & 292.1    & A\\
14897 & 2013-08-07    16:59:23 & 50.6 & 266.5801 & -28.8682 & 273.8  & A,Q\\
17236 & 2015-04-25    14:10:43 & 79.0 & 266.5756 & -28.9138 & 85.2   & Q\\
17237 & 2016-05-18    05:20:13 & 20.8 & 266.5619 & -28.9116 & 77.2   & Q\\
17238 & 2017-07-17    18:45:57 & 65.2 & 266.5581 & -28.9189 & 278.2  & Q\\
17239 & 2015-08-19    17:23:06 & 79.0 & 266.5276 & -28.8906 & 272.1  & A,Q\\
17240 & 2016-07-24    05:51:20 & 74.7 & 266.5376 & -28.8980 & 276.8  & Q\\
17241 & 2017-10-02    18:55:02 & 24.8 & 266.5264 & -28.8895 & 267.2  & A,Q\\
18324 & 2016-07-15    06:33:43 & 1.9 & 266.5392 & -28.7562 & 279.7   & A,Q\\
18852 & 2016-05-18    20:50:51 & 52.4 & 266.5622 & -28.9123 & 77.2   & Q\\
20118 & 2017-07-23    01:21:05 & 13.9 & 266.5579 & -28.9190 & 278.2  & Q\\
20807 & 2017-10-05    14:01:00 & 27.7 & 266.5263 & -28.8894 & 267.2  & Q\\
20808 & 2017-10-08    15:38:21 & 26.7 & 266.5265 & -28.8897 & 267.2  & A,Q\\
24823 & 2021-06-07    01:25:27 & 9.0 & 266.5963 & -28.8674 & 65.2    & A,Q\\
25059 & 2021-06-07    16:05:38 & 20.8 & 266.5964 & -28.8680 & 65.2   & A,Q\\
24362 & 2022-02-02    13:36:10 & 13.9 & 266.5450 & -28.8827 & 93.6   & A,Q\\
26294 & 2022-02-04    05:48:53 & 15.0 & 266.5450 & -28.8821 & 90.5   & A,Q\\
24367 & 2022-02-05    07:49:12 & 17.7 & 266.5453 & -28.8823 & 88.5   & A,Q\\
26295 & 2022-02-05    18:19:32 & 11.9 & 266.5451 & -28.8822 & 87.9   & A,Q\\
24373 & 2022-02-26    22:53:25 & 27.7 & 266.5774 & -28.9209 & 81.2   & A,Q\\
24820 & 2022-03-09    14:33:56 & 12.9 & 266.5768 & -28.9218 & 80.2   & A,Q\\
24358 & 2022-03-10    11:05:57 & 18.6 & 266.5767 & -28.9220 & 80.2   & A,Q\\
26353 & 2022-03-11    20:23:08 & 10.1 & 266.5771 & -28.9221 & 80.2   & A,Q\\
26355 & 2022-03-12    06:46:50 & 16.7 & 266.5767 & -28.9219 & 80.2   & A,Q\\
23641 & 2022-04-08    12:51:30 & 27.7 & 266.5453 & -28.8832 & 86.6   & A,Q\\
24365 & 2022-04-10    05:57:20 & 29.3 & 266.5098 & -28.8479 & 86.4   & A,Q\\
24370 & 2022-04-17    20:45:22 & 29.5 & 266.5106 & -28.8475 & 85.5   & A,Q\\
24821 & 2022-06-08    01:04:09 & 27.2 & 266.4927 & -28.8847 & 45.2   & A,Q\\
26302 & 2022-06-11    06:43:11 & 19.1 & 266.4929 & -28.8852 & 46.2   & A,Q\\
24360 & 2022-06-19    21:30:30 & 14.9 & 266.5921 & -28.8660 & 342.7  & A,Q\\
26438 & 2022-06-20    05:54:23 & 14.9 & 266.5921 & -28.8660 & 342.7  & A,Q\\
24363 & 2022-06-26    06:45:35 & 27.7 & 266.4939 & -28.8678 & 289.2  & A,Q\\
24376 & 2022-07-01    23:43:51 & 18.8 & 266.5138 & -28.9287 & 284.2  & A,Q\\
26445 & 2022-07-02    11:21:08 & 10.0 & 266.5142 & -28.9294 & 284.2  & A,Q\\
24369 & 2022-07-06    11:16:01 & 15.9 & 266.5680 & -28.8469 & 285.7  & A,Q\\
26448 & 2022-07-07    04:35:42 & 16.4 & 266.5685 & -28.8476 & 285.7  & A,Q\\
26303 & 2022-07-20    05:52:01 & 15.8 & 266.5088 & -28.9257 & 278.3  & A,Q\\
24368 & 2022-07-20    19:17:22 & 27.2 & 266.4986 & -28.8598 & 277.9  & A,Q\\
24822 & 2022-07-30    00:36:42 & 17.7 & 266.5825 & -28.9180 & 275.5  & A,Q\\
24372 & 2022-08-02    13:46:45 & 32.7 & 266.5413 & -28.8884 & 274.7  & A,Q\\
26304 & 2022-08-06    20:42:48 & 11.9 & 266.5813 & -28.9192 & 274.0  & A,Q\\
24364 & 2022-08-07    00:25:20 & 10.5 & 266.5767 & -28.8524 & 273.9  & A,Q\\
24818 & 2022-08-07    03:32:11 & 13.8 & 266.5055 & -28.9230 & 273.9  & A,Q\\
24375 & 2022-08-07    14:45:41 & 27.5 & 266.5816 & -28.9193 & 273.9  & A,Q\\
24750 & 2022-08-30    16:00:20 & 10.9 & 266.5409 & -28.8882 & 270.9  & A,Q\\
24819 & 2022-09-03    15:07:37 & 26.9 & 266.5411 & -28.8880 & 270.5  & A,Q\\
24816 & 2022-09-13    06:22:20 & 29.7 & 266.5791 & -28.8546 & 269.5  & A,Q\\
24359 & 2022-09-15    06:24:12 & 29.7 & 266.5796 & -28.8555 & 269.3  & A,Q\\
24371 & 2022-09-17    02:28:56 & 29.0 & 266.5023 & -28.9214 & 269.1  & A,Q\\
24361 & 2022-09-23    04:35:09 & 29.7 & 266.4962 & -28.9153 & 260.2  & A,Q\\
24374 & 2022-09-26    03:33:08 & 27.5 & 266.5797 & -28.8554 & 268.3  & A,Q\\
27256 & 2022-09-29    23:19:43 & 19.8 & 266.5404 & -28.8877 & 268.0  & A,Q\\
24815 & 2022-10-01    02:51:19 & 27.7 & 266.5026 & -28.8544 & 270.2  & A,Q\\
24817 & 2022-10-02    23:00:43 & 28.0 & 266.5804 & -28.9202 & 272.2  & A,Q\\
24366 & 2022-10-09    12:44:53 & 14.7 & 266.5036 & -28.9226 & 270.2  & A,Q\\
27463 & 2022-10-09    22:12:42 & 14.9 & 266.5036 & -28.9227 & 270.2  & A,Q\\
\hline
\end{longtable}

}

\twocolumn





\bsp	
\label{lastpage}
\end{document}